\documentclass[aps,prb,floatfix,epsfig,twocolumn,showpacs,preprintnumbers]{revtex4}
\usepackage{amsfonts}
\usepackage{amssymb}
\usepackage{graphicx}
\usepackage{amsmath}

\begin{document}

\title{Electronic structure of Na, K, Si, and LiF from self-consistent solution of Hedin's equations including vertex corrections}
\author{Andrey L. Kutepov\footnote{e-mail: kutepov@physics.rutgers.edu}}
\affiliation{Department of Physics and Astronomy, Rutgers University, Piscataway, NJ 08856}

\begin{abstract}
A few self-consistent schemes to solve the Hedin equations are presented. They include vertex corrections of different complexity. Commonly used quasiparticle approximation for the Green function and static approximation for the screened interaction are avoided altogether. Using alkali
metals Na and K as well as semiconductor Si and wide gap insulator LiF as examples, it is shown that both the vertex corrections in the
polarizability P and in the self energy $\Sigma$ are important. Particularly, vertex corrections in $\Sigma$ with proper treatment of frequency dependence of the screened interaction always reduce calculated band widths/gaps, improving the agreement with experiment. The complexity of the vertex included in P and in
$\Sigma$ can be different. Whereas in the case of polarizability one generally has to solve the Bethe-Salpeter equation for the corresponding vertex function,
it is enough (for the materials in this study) to include the vertex of the first order in the self energy. The calculations with appropriate
vertices show remarkable improvement in the calculated band widths and band gaps as compared to the self-consistent GW approximation as well as to the self-consistent quasiparticle GW approximation.
\end{abstract}

\maketitle

\section*{Introduction}
\label{intro}

Since its first implementation by Hybertsen and Louie\cite{prb_34_5390} and by Godby et al\cite{prb_37_10159} the so called  $G_{0}W_{0}$ method (with $G_{0}$ being the Green function
in the local density approximation (LDA) for the density functional theory, and $W_{0}$ being the screened interaction in the random phase
approximation (RPA)) has become a method of choice for relatively inexpensive and accurate calculations of the electronic structure of weakly
correlated materials.\cite{cpc_182_2029,cpc_183_1269,cpc_184_348,jcm_26_363202,njp_14_023006,prb_66_125101,prb_74_035101,prb_81_125102,
prb_87_155148,prb_87_235102,prb_87_235132,prb_90_075125,prb_91_155141,prb_93_115203,prb_94_035103,prb_94_035118,
prl_101_106404,prl_113_076402,prl_88_016403,prb_88_075117} As a disadvantage of the approach one can point out its dependence on the starting point. Electronic spectra obtained in LDA
should already be sufficiently accurate in order to ensure that $G_{0}W_{0}$ provides results close to the experiment. Obviously, it is not always
the case. One of the remedies is to switch from the LDA to another starting point which suits better for the specific material. For example Jiang
et al.\cite{prb_86_125115} used LDA+U method\cite{prb_48_16929} to study the electronic structure of lanthanide oxides. By adjusting the U
parameter one can construct the LDA+U spectra in decent agreement with experiment and correspondingly the $G_{0}W_{0}$ approach performed on top
of LDA+U may work pretty well. One can use other starting points together with the $G_{0}W_{0}$ approach: exact exchange approximation (EXX)\cite{pss_245_929,njp_7_126} or hybrid functional\cite{prb_76_115109,pss_246_1877}. Generally one can say that the success of the G$_{0}$W$_{0}$ approach is based on the cancellation of error stemming for the lack of self-consistency on the one hand and the absense of the vertex corrections on the other hand. Whereas different starting points followed by $G_{0}W_{0}$ iteration may reproduce the experimental spectra with
good accuracy for variety of materials, the approach can hardly be considered as a satisfactory one.

A logical way to eliminate the dependence on the starting point is to perform $GW$ calculations self consistently (sc). However, fully self
consistent $GW$ approach without vertex corrections has certain theoretical problems\cite{tcc_347_99} and corresponding calculations overestimate band gaps in semiconductors and insulators, and band widths in
metals.\cite{prb_85_155129} It seems to be better justified for applications in the physics of atoms and molecules, as one can judge from the noticeable progress in the field.\cite{jcp_141_084108,jctc_9_232,jctc_10_3934,epl_76_298,prb_75_205129,prb_81_085103,prb_86_081102,prb_88_075105,
prb_89_155417,prb_91_125135,prb_91_205111,prb_93_121115,prl_110_146403} In the physics of solids, however, considerable requirements of the fully scGW method to the computer resources as well as intrinsic problems of the method itself\cite{tcc_347_99} have made it quite common to use partially scGW schemes. Among such partially scGW approaches one can mention GW$_{0}$ scheme\cite{prb_92_041115,prl_99_246403,prb_93_075205,prb_93_115203,prapp_6_014009} where W is fixed at RPA level (usually calculated with LDA Green's function) and only G is iterated till convergence. Another popular approach is the so called energy-only self-consistent GW\cite{prb_34_5390,prb_83_115103,prb_83_115123,prl_59_819,prb_66_195215,prb_71_045207,prb_75_235102} where one-electron wave functions are fixed (again, usually at LDA level) and only one-electron energies are renewed till consistency. The success of these partially sc schemes is based on the same cancellation of errors as in the case of G$_{0}$W$_{0}$. Partial sc usually makes the spectral features (band widths/gaps) a little wider and, thus, often improves the agreement with experiment. Authors of the Refs.[\onlinecite{prl_89_126401}] and [\onlinecite{prb_93_125210}] propose to apply diagonal (in LDA band states basis) approximation for the self energy and Green's function which makes the calculations much faster. In this case the success is based on the cancellation of error stemming from the neglect of non-diagonal terms in G and $\Sigma$ on the one hand and the neglect of vertex corrections on the other hand.

Considerable progress has been made by Kotani et al.\cite{prb_76_165106} in their QSGW approach which essentially is
equivalent to the fully scGW method but with special (quasi-particle (QP)) construction for the Green function, which replaces the need to solve
the Dyson equation. The success of QSGW method relies on the fact that QP approximation cancels out in considerable degree the error associated
with the absence of higher order diagrams in the self energy $\Sigma$ and the polarizability $P$, as it has been explained in
Ref.[\onlinecite{prb_76_165106}] in terms of Z-factor cancelation. QSGW approach is computationally more expensive than $G_{0}W_{0}$ but it doesn't
depend on a starting point. It usually gives the results similar to the LDA-based $G_{0}W_{0}$ results for simple metals and semiconductors, but
often shows improvements for the materials where LDA doesn't provide a good starting point for the $G_{0}W_{0}$ iteration (NiO is a good example,
as it has been shown by Faleev et al.\cite{prl_93_126406}).

Presently, QSGW approach is a very popular ab-initio method which provides reasonable one electron spectra for a wide class of materials.\cite{prb_76_165126,prl_93_126406,prl_97_267601,prl_96_226402,prb_81_045203,prb_81_125201,jcm_20_295214,
prb_81_245120,prb_84_205205,prb_84_165204,prl_99_246403,prb_87_155147}
However, even for relatively weakly correlated materials, there is still enough room for improvements. Looking at the results obtained with the QSGW
method\cite{prb_76_165106,prl_99_246403,prb_92_041115} one can conclude that calculated band gaps are overestimated by about 5-15\% for
sp-semiconductots and insulators. For the materials with d- and f-electrons (SrTiO$_{3}$, TiO$_{2}$, CeO$_{2}$) the error grows up to about
25\%.\cite{prl_96_226402} Similar error has been found in the calculated exchange splitting in Gadolinium\cite{prl_96_226402}, whereas the
calculated exchange splitting in Nickel is almost twice too large as compared to the experimental one.\cite{prl_96_226402} Besides, with QP
construction for the Green function the method is not diagrammatic anymore, which renders its improvement more complicated.

An alternative way to improve the accuracy of the scGW method is to include skeleton diagrams of higher order (vertex corrections) in the self energy and the
polarizability. However, direct diagrammatic extensions of this kind represent an extremely difficult problem in practice and, as a result, were not explored actively for solids. Ummels et al.\cite{prb_57_11962} have applied first order
vertex corrections to $P$ and $\Sigma$ combined with second order self-consistence diagrams for
silicon and diamond. Calculations have been performed with LDA Green's function and within plasmon pole approximation\cite{prb_47_15931}. It has been shown that vertex corrections and self-consistence diagrams cancel out to a high degree (especially the correction to P) which can be considered as a
justification for the one-shot $G_{0}W_{0}$ approach. Bechstedt et al\cite{prl_78_1528} iterated the Dyson equation for G and the Bethe-Salpeter equation for the irreducible polarizability simultaneously. Certain approximations (such as keeping only diagonal terms in bloch integrals and neglect of the local field effect) have been made in the study. The principal conclusion of the work is that vertex correction in polarizability widely compensates the GW quasiparticle peaks renormalization, which can be considered as a support in favor of the QSGW approximation.

Considerable progress has been achieved, however, in studying the effect of vertex corrections following the ideas borrowed from the Time Dependent Density Functional Theory (TDDFT)\cite{prl_45_204,prl_52_997,prl_55_2850}, where the central role is played by the so called exchange-correlation kernel $f_{xc}$. The research along this line began in Refs.[\onlinecite{prb_49_8024,prl_62_2718,prb_56_12832,prb_76_155106}] where LDA-based 2-point vertex function was proposed. Model exchange-correlation kernels have also been introduced\cite{prb_92_041115,prl_107_186401,prb_87_205143,prb_69_155112,prb_72_125203} with improved (as compared to LDA-derived kernel) properties. A very successful approach has been developed which recasts diagrammatically obtained polarizability (usually of low order) into an effective exchange-correlation kernel $f_{xc}$.\cite{prl_88_066404,prb_68_165108,prl_91_056402,prb_70_081103,prl_94_186402,rpp_70_357} The kernel $f_{xc}$ is a two-point object (as opposed to the many-body kernel which is a four-point object). So the above recasting brings in a great
efficiency. Shishkin et al.\cite{prl_99_246403} have applied this approach to calculate the band gaps for a wide class of materials. The results obtained in Ref.\onlinecite{prl_99_246403} look promising. However, there were many simplifications involved in the
calculations. First of all, the vertex correction has been included only in the polarizability, but not in the self energy. Second, it was static,
i.e. W in the diagrams has been approximated by its value at zero frequency. But may be most important of all is the fact that authors applied the vertex
correction combined with quasi-particle self-consistence. The problem with this kind of approach is that the quasi-particle approximation
itself can be considered as an effective vertex correction (due to Z-factor cancellation). If one applies the same arguments, as the authors of
Ref.[\onlinecite{prb_76_165106}] did, to the approach which combines the QSGW and the vertex corrections one will realize that there is a double
counting. The problems of combining the QSGW approach with vertex corrections have been studied for the two-site Hubbard model
recently.\cite{jcm_27_315603} Based on the above consideration, one can speculate that the static approximation for $W$ was actually needed to cancel
out the error stemming from that double counting, because zero-frequency interaction is well enough screened and, correspondingly, its effect is
much weaker than it would be had the authors of Ref.\onlinecite{prl_99_246403} applied full frequency-dependent interaction. As for the absence of the vertex correction in the self
energy, authors say that their inclusion "turned out to be numerically rather unstable and tended to bring the band gaps back to those obtained
without vertex corrections", which can also be considered as a sign of inherent problems with the approach. A similar approach (combination of
QSGW with static $f_{xc}$) has been used recently by Gruneis et al.\cite{prl_112_096401} to study the ionization potentials and band gaps of
solids. In addition, authors of Ref.[\onlinecite{prl_112_096401}] have considered the correction to the self energy of the second order, but
again, evaluated with static interaction. Their observation was that vertex correction in the self energy actually increases the band gaps, making
them worse than the ones with the vertex correction only in the polarizability.

In the present work the above simplifications in dealing with the vertex corrections are avoided. The approach is based on
the Hedin exact theory\cite{pr_139_A796} and approximations are introduced purely diagrammatically, without connection with TDDFT. Also, there is no quasi-particle
approximation involved. Instead the Green function is renewed on every iteration from Dyson's equation. All diagrams take into account full
frequency-dependence of the screened interaction, which also is updated on every iteration. Third, the vertex corrections are studied for both the
polarizability and the self energy.

The principal goal of this study is to elucidate the effect of vertex corrections in fully self-consistent calculations. To make this research as clean as possible, one has to avoid the schemes which are based on the cancellation of errors. This makes the direct comparison of the methods being developed in this work with previous studies (based on G$_{0}$W$_{0}$, GW$_{0}$, QSGW, QSGW + vertex evaluated with static W) not very useful for answering the main question of this research. Comparison with earlier studies is very useful, however, to check the accuracy of numerical implementation of the code.

The paper begins with a formal presentation of Hedin's equations (subsection \ref{Hed_eq}). The self-consistent schemes of solving them together
with numerical approximations comprise the subsections \ref{vrt_appr}, \ref{def_sc}, and \ref{num_appr}. Section \ref{res} provides the
results obtained and a discussion. The conclusions are given afterwords. Finally, the details of the practical solution of Hedin's equations for
solids are presented in the Appendix.

\section{Method}
\label{meth}

\subsection{Hedin's equations} \label{Hed_eq}

The approach which is used in this work is based on the Hedin equations.\cite{pr_139_A796} For convenience, we remind the reader about how Hedin's
equations could be solved self-consistently in practice. Matsubara's formalism is used throughout the work.

Suppose one has a certain initial approach for the green function $G$ and the screened interaction $W$. Then one calculates the following quantities:

three-point vertex function from the Bethe-Salpeter equation

\begin{align}\label{Vert_0}
\Gamma^{\alpha}(123)&=\delta(12)\delta(13)\nonumber\\&+\sum_{\beta}\frac{\delta \Sigma^{\alpha}(12)}{\delta
G^{\beta}(45)}G^{\beta}(46)\Gamma^{\beta}(673) G^{\beta}(75),
\end{align}
where $\alpha$ and $\beta$ are spin indexes, and the digits in the brackets represent space-Matsubara's time arguments,

polarizability

\begin{equation}\label{def_pol1}
P(12)=\sum_{\alpha}G^{\alpha}(13)\Gamma^{\alpha}(342)G^{\alpha}(41),
\end{equation}

screened interaction

\begin{align}\label{def_W2}
W(12)=V(12) +V(13)P(34)W(42),
\end{align}

and the self energy

\begin{equation}\label{def_M8}
\Sigma^{\alpha}(12)= - G^{\alpha}(14)\Gamma^{\alpha}(425)W(51).
\end{equation}

In the equation (\ref{def_W2}) V stands for the bare Coulomb interaction.
New approximation for the Green function is obtained from Dyson's equation

\begin{align}\label{D4}
G^{\alpha}(12)=G_{0}^{\alpha}(12) +G_{0}^{\alpha}(13)\Sigma^{\alpha}(34)G^{\alpha}(42),
\end{align}
where $G_{0}$ is the Green function in Hartree approximation. Eqn. (\ref{Vert_0}-\ref{D4}) comprise one iteration. If convergence is not yet reached one can go back to the equation (\ref{Vert_0}) to
start the next iteration with renewed $G$ and $W$.

The system of Hedin's equations formally is exact, but one has to introduce certain approximations when solving (\ref{Vert_0}) for the vertex function $\Gamma^{\alpha}(123)$ in order to make the
system manageable in practice.

\subsection{Approximations for the vertex function} \label{vrt_appr}

\begin{figure}[b]
\includegraphics[width=7.0 cm]{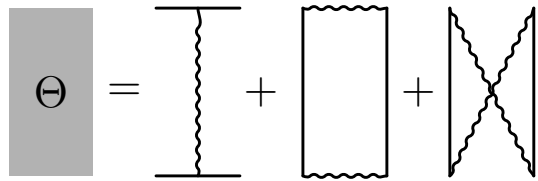}
\caption{The GW approximation for the irreducible 4-point kernel $\Theta$. Direct lines represent Green's function and wavy lines represent screened interaction W.} \label{theta_appr}
\end{figure}

A convenient way to generate approximations for the vertex $\Gamma$ is to calculate the kernel $\Theta=\frac{\delta \Sigma}{\delta
G}$ in Eqn.(\ref{Vert_0}) using a diagrammatic representation of the self energy up to a specific order in the screened interaction $W$. The simplest non-trivial approach in this case is to use the famous $GW$ approximation
($\Sigma=GW$) where $W$ is obtained from the polarizability ($W=V+VPW$) which in turn is represented by the one-loop approximation ($P=GG$).
Adapting this approach one gets for the kernel:

\begin{align}\label{dsig_dg0}
&\frac{\delta \Sigma^{\alpha}(12)}{\delta G^{\beta}(34)}=-\delta_{\alpha\beta}\delta(13)\delta(24)W(21)\nonumber\\&
-G^{\alpha}(12)G^{\beta}(43)\left[W(23)W(41)+W(24)W(31)\right],
\end{align}
which is shown diagrammatically in Fig.\ref{theta_appr}.

Approximation (\ref{dsig_dg0}) results in the following equation for the vertex function:

\begin{align}\label{Vert_0_1}
&\Gamma^{\alpha}(123)=\delta(12)\delta(13)-W(21)G^{\alpha}(14)\Gamma^{\alpha}(453)
G^{\alpha}(52)\nonumber\\&-G^{\alpha}(12)\sum_{\beta}G^{\beta}(54)[W(24)W(51)+W(25)W(41)]\nonumber\\&\times G^{\beta}(46)\Gamma^{\beta}(673)
G^{\beta}(75).
\end{align}

\begin{figure}[t]
\includegraphics[width=5.0 cm]{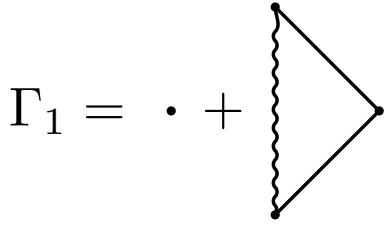}
\caption{First order approximation for the 3-point vertex function.}
\label{gamma_1}
\end{figure}

\begin{figure}[b]
\includegraphics[width=8.0 cm]{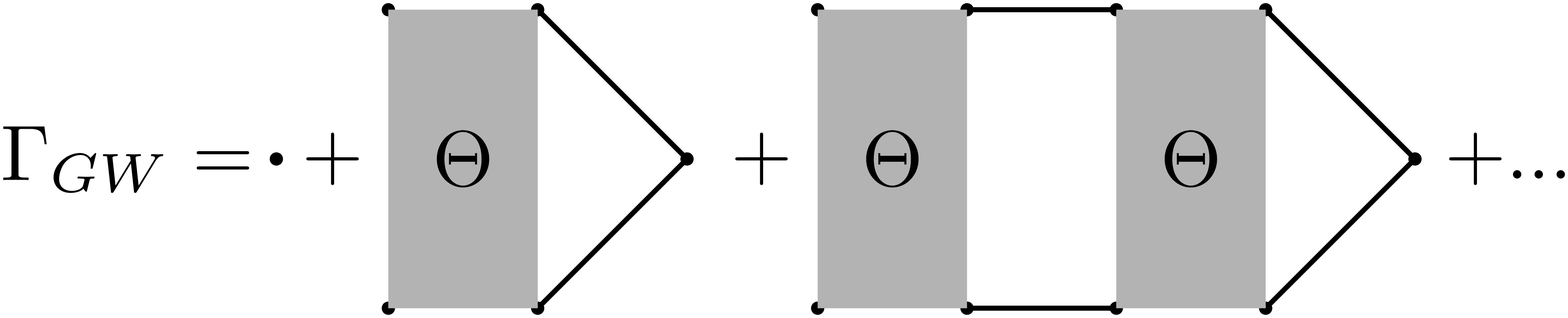}
\caption{Ladder sequence for the 3-point vertex function with
$\Theta$ as the rung of the ladder.} \label{gamma_theta}
\end{figure}

\begin{figure}[t]
\includegraphics[width=8.0 cm]{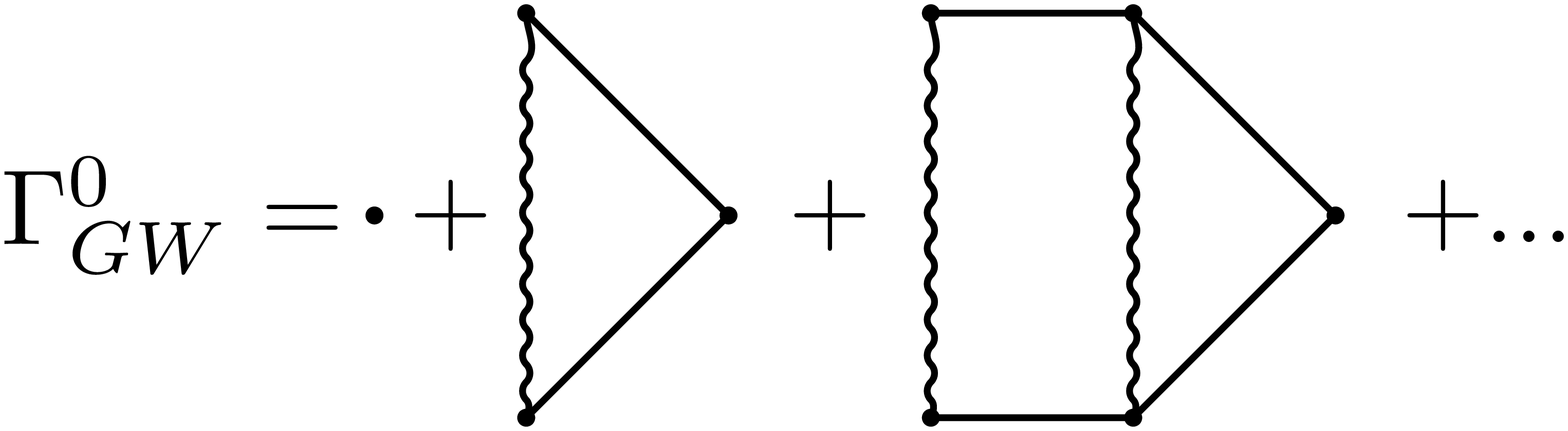}
\caption{Ladder sequence for the 3-point vertex function with $W$ as
the rung of the ladder.} \label{gamma_w}
\end{figure}

It is convenient to split the vertex into a trivial part and a correction ($\Gamma=1+\triangle\Gamma$). In this case one obtains an equation for the correction which might
be solved iteratively:

\begin{align}\label{d_Vert1}
\triangle\Gamma^{\alpha}(123)=&-W(2,1)G^{\alpha}(13)G^{\alpha}(32)\nonumber\\&-W(2,1)G^{\alpha}(14)\triangle\Gamma^{\alpha}(453)
G^{\alpha}(52)\nonumber\\-G^{\alpha}(12)\sum_{\beta}&G^{\beta}(54)[W(24)W(51)+W(25)W(41)]\nonumber\\ \times
[G^{\beta}(43)&G^{\beta}(35)+G^{\beta}(46)\triangle\Gamma^{\beta}(673) G^{\beta}(75)].
\end{align}

In this work the following non-trivial approximations for the vertex are used: i) first order approximation ($\Gamma_{1}$) is obtained when one
keeps only the first term on the right hand side of (\ref{d_Vert1}) (schematically $\Delta\Gamma_{1}=-WGG$), ii) the vertex in "GW" approximation
($\Gamma_{GW}$) when all terms on the rhs of (\ref{d_Vert1}) are kept intact, and iii) the vertex $\Gamma^{0}_{GW}$, which is similar to the
approximation $\Gamma_{GW}$, but corresponds to an additional approximation where one neglects the diagrams with possible spin flips (i.e. the
terms with $\sum_{\beta}$ in (\ref{d_Vert1}) are not included). Diagrammatic representations of the approximations i)-iii) are shown in Figures \ref{gamma_1}, \ref{gamma_theta}, and \ref{gamma_w} correspondingly. The abbreviation $\Gamma_{GW}$ is particularly meaningful when the corresponding vertex is calculated
with $G$ and $W$ from scGW calculation. The polarizability evaluated with this vertex (and with $G$ in (\ref{def_pol1}) also taken from scGW) is
"physical" in a sense that it is an exact functional derivative of the electronic density (calculated in scGW approximation) with respect to the
total electric field (external plus induced) and, as a result, respects the charge preservation. In the present work another variant of
$\Gamma_{GW}$ is used - with $G$ and $W$ being fully self-consistent (with vertex corrections included). In this case the corresponding
polarizability is no more physical because the self energy and the polarizability include more diagrams than the approximation ($\Sigma=GW$, $P=GG$) assumed here in the Bethe-Salpeter equation. Thus, in fully self-consistent calculations one has to trade between the improvements in spectra resulting from higher order diagrams on the one hand and the degree of charge preservation on the other hand.

In this study the vertex $\Gamma_{GW}$ is calculated from Eq.(\ref{d_Vert1}) iteratively, i.e. the calculation of the vertex function is achieved
through a "small" loop of iterations as compared to the "big" loop of iterations of the self-consistent scheme depicted in the
Eqs.(\ref{Vert_0}-\ref{D4}). Corresponding steps of the "small" loop of iterations are sketched below. Full details are given in Appendix.

To simplify the formulae the following notations are introduced

\begin{align}\label{K0_def}
K^{0\alpha}(123)=-G^{\alpha}(13)G^{\alpha}(32),
\end{align}

\begin{align}\label{dK_def}
\triangle K^{\alpha}(123)=-G^{\alpha}(14)\triangle\Gamma^{\alpha}(453) G^{\alpha}(52),
\end{align}

\begin{align}\label{K_def}
K^{\alpha}(123)=K^{0\alpha}(123)+\triangle K^{\alpha}(123),
\end{align}
so that the equation (\ref{d_Vert1}) for the correction to the vertex takes the following form

\begin{align}\label{d_Vert2}
&\triangle\Gamma_{\alpha}(123)=W(21)K^{\alpha}(123)+G^{\alpha}(12)\nonumber\\&\times
\sum_{\beta}W(24)\big[G^{\beta}(54)K^{\beta}(453)+G^{\beta}(45)K^{\beta}(543)\big]W(51).
\end{align}

Introducing yet more of notations

\begin{align}\label{Q_def}
Q(123)=\sum_{\beta}\big[G^{\beta}(21)K^{\beta}(123)+G^{\beta}(12)K^{\beta}(213)\big],
\end{align}
and
\begin{align}\label{T_def}
T(213)=W(24)Q(453)W(51),
\end{align}
one reduces the equation for the vertex correction to a formally very simple form

\begin{align}\label{d_Vert3}
\triangle\Gamma_{\alpha}(123)=W(21)K^{\alpha}(123)+G^{\alpha}(12)T(213).
\end{align}

The iterations for the $\Gamma_{GW}$ are performed as the following. One takes $K=K^{0}$ (Eqn. \ref{K0_def}) as an initial approach, then calculates
$Q$ (Eqn. \ref{Q_def}), $T$ (Eqn. \ref{T_def}), and $\triangle \Gamma$ (Eqn. \ref{d_Vert3}). Then a correction to $K^{0}$ (Eqn.
\ref{dK_def}) is evaluated and the process is repeated with a new $K=K^{0}+\triangle K$. The iterations for the $\Gamma^{0}_{GW}$ are simpler. They follow the
same scheme but without $Q$ and $T$ evaluation. Finally, the approximation $\Gamma_{1}$ is obtained with just one step:
$\triangle\Gamma_{1}=WK^{0}$.

Some of the above equations are easier to handle in the reciprocal space (band representation) and frequency, whereas other are simpler in the
real space and imaginary time representation. So one switches from one to another representation and back on every iteration. The details about how
it is done can be found in Appendix.

\subsection{Definitions of self-consistent schemes} \label{def_sc}

Having defined the approximations for the vertex function one can proceed with the construction of iterative schemes of solving the Hedin equations
(\ref{Vert_0}-\ref{D4}). The schemes differ by which approximation for the vertex function is used in the expression for the polarizability
(\ref{def_pol1}) and in the expression for the self energy (\ref{def_M8}). In this work seven sc schemes are studied. They have been collected in
Table \ref{sc_schemes} which explains their diagrammatic representations.

\begin{table}[b]
\caption{Diagrammatic representations of the polarizability and the self energy in sc schemes of solving the Hedin equations. Arguments in square
brackets specify $G$ and $W$ which are used to evaluate the vertex function. Other details are explained in the main text.} \label{sc_schemes}
\begin{center}
\begin{tabular}{@{}c c c} Scheme  & $P$ & $\Sigma$\\
\hline
A & $GG$ & $GW$ \\
B & $G\Gamma_{1}[G;W]G$ & $G\Gamma_{1}[G;W]W$ \\
C & $\underline{G}\Gamma_{GW}[\underline{G};\underline{W}]\underline{G}$ & $G\overline{W}$ \\
D & $\underline{G}\Gamma_{GW}[\underline{G};\underline{W}]\underline{G}$ & $G\Gamma_{1}[G;\overline{W}]\overline{W}$ \\
E & $G\Gamma_{GW}[G;W]G$ & $G\Gamma_{1}[G;W]W$ \\
F & $G\Gamma_{GW}[G;W]G$ & $G\Gamma_{GW}[G;W]W$ \\
G & $G\Gamma^{0}_{GW}[G;W]G$ & $G\Gamma_{1}[G;W]W$
\end{tabular}
\end{center}
\end{table}

Scheme A is scGW approach.  It is conserving in Baym-Kadanoff definition,\cite{pr_124_287} but generally its accuracy is poor when one considers spectral properties of solids.\cite{prb_80_041103,prb_85_155129,prl_89_126402} Another
conserving sc scheme is the scheme B. It uses the same first order vertex in both $P$ and $\Sigma$. Scheme C is based on "physical" polarizability
as it was explained in the subsection \ref{vrt_appr}. We perform scGW calculation first. Underlined $\underline{G}$ and $\underline{W}$ in Table \ref{sc_schemes}
mean that the corresponding quantities are taken from scGW run. Then the vertex $\Gamma_{GW}[\underline{G};\underline{W}]$ is evaluated and it is
used to calculate the polarizability and the corresponding screened interaction $\overline{W}$. We use a bar above the $W$ to indicate that this
quantity is evaluated using $\underline{G}$ and $\underline{W}$ from scGW calculation, but it is not equal to $\underline{W}$ because it includes
vertex corrections through the polarizability. This $\overline{W}$ is fixed (in the scheme C) during the following iterations where only the self
energy $\Sigma=G\overline{W}$ and $G$ are renewed. So, the scheme C doesn't include vertex in $\Sigma$ explicitly but only through $\overline{W}$.
The scheme D is similar to the scheme C. It also is based on the physical polarizability but it uses the first order vertex in the self energy
explicitly (skeleton diagram). In the scheme D the screened interaction $W$ is fixed at the same level as in the scheme C, but the final iterations involve the
renewal of not only $G$ and $\Sigma$, but also $\Gamma_{1}$. The schemes E and F are fully self-consistent (both $G$ and $W$ are renewed on every iteration till
the end). They differ only in the diagrammatic representation of the self energy. As it was pointed out in the previous section, the schemes E and F do not preserve the charge exactly with the scheme F being potentially more problematic because the imbalance between the kernel of the Bethe-Salpeter equation and the diagrammatic representation of $\Sigma$ in the scheme F is larger. Scheme G is similar to the scheme E, but with simplified
Bethe-Salpeter equation for the corresponding vertex $\Gamma^{0}_{GW}$ (the diagrams with spin-flips are neglected in the kernel of the Bethe-Salpeter
equation).

Table \ref{schemes_prop} collects the features of the above schemes for convenience.

\begin{table}[t]
\caption{Properties of the sc schemes studied in this work.} \label{schemes_prop}
\begin{center}
\begin{tabular}{@{}c c c c c c c c} Property  & A & B & C & D & E & F & G\\
\hline
Conserving & yes & yes & no & no & no & no & no \\
$P$ is physical & no & no & yes & yes & no & no & no \\
Same vertex in $P$ and $\Sigma$ & yes & yes & no & no & no & yes & no\\
Self-consistency & full & full & partial & partial & full & full & full
\end{tabular}
\end{center}
\end{table}

\section{Numerical approximations} \label{num_appr}

Vertex corrected calculations generally are very computationally expensive as compared to scGW calculations. If one implements higher order diagrams using the same basis set, and the same number of k-points as for the evaluation of GW diagram, the evaluation of them (higher order diagrams) will be prohibitively expensive. However, what makes this kind of calculations feasible is the fact that vertex part is effective on the lower energy scale (i.e. only near Fermi level) as compared to the GW part. This fact allows us to use smaller basis sets in the vertex part, which in its own turn allows to use coarser time/frequency meshes to represent vertex-dependent functions. Also, the diagrams beyond GW are generally more localized in real space (see discussions in Refs.[\onlinecite{plb_29_632,prl_96_226403}]), which allows one to use coarser k-mesh for their evaluation.

\begin{figure}[t]
\centering
\includegraphics[width=8.5 cm]{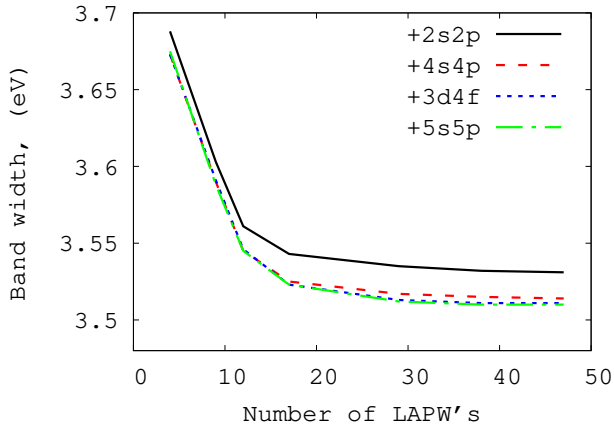}
\caption{(Color online) Convergence of the band width (Na) with respect to the size of LAPW basis set. Different lines correspond to the addition of more and more local orbitals (LO) to the pure LAPW basis as indicated in legends.} \label{lapw_na}
\end{figure}

\begin{figure}[t]
\centering
\includegraphics[width=8.5 cm]{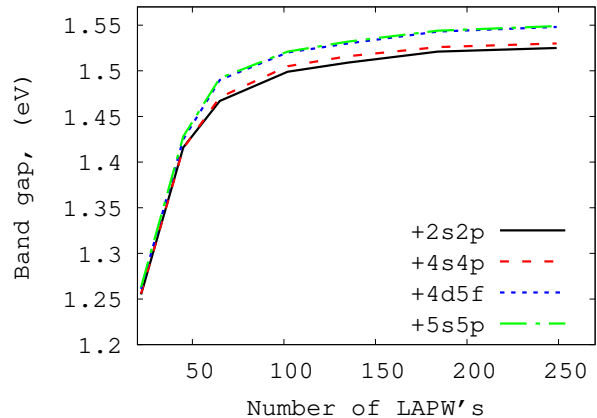}
\caption{(Color online) Convergence of the band gap (Si) with respect to the size of LAPW basis set. Different lines correspond to the addition of more and more local orbitals (LO) to the pure LAPW basis as indicated in legends.} \label{lapw_si}
\end{figure}

In this work only one of the above three possible optimizations has been explored. Namely, the number of bands which were used to represent Green's function and self energy in GW part and in the vertex part were different. Thus, the tests of convergence with respect to the basis set size have been performed separately for the GW part and for the vertex part. These tests and all other convergence tests have been conducted for one metal (Na) and for one material with a gap (Si). Figures \ref{lapw_na} and \ref{lapw_si} show the convergence of the band width (Na) and the band gap (Si) in scGW (scheme A). In this work the number of band states used as a basis set for GW part was equal to the size of FLAPW+LO basis set. So the figures \ref{lapw_na} and \ref{lapw_si} show essentially the convergence of scGW results with respect to the number of linearized augmented plane waves and local orbitals. k-meshes $12\times 12\times 12$ and $8\times 8\times 8$ have been used for Na and Si correspondingly in  getting the data for plots.

As one can see the convergence is very fast for Na, but slow enough for Si. However it posed no problem for the present research as the really time consuming part was the vertex part. Thus, in all presented below results the FLAPW+LO basis set in GW part was well converged for all four materials. The size of product basis set (PB) for the GW part was not independent and was adjusted for every change in the size of FLAPW+LO basis set. The criterium for this adjustment was the requirement that the convolution of G and $\Sigma$ (they are represented in band states basis) and the convolution of P and W (they are represented in PB) were the same within given tolerance ($10^{-4}$ Ry in this work).

\begin{figure}[b]
\centering
\includegraphics[width=8.5 cm]{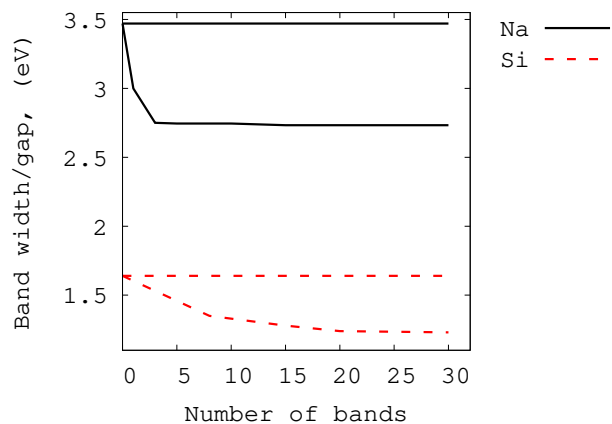}
\caption{(Color online) Convergence of the band width (Na) and $\Gamma-X$ band gap (Si) with respect to the number of band states included in the vertex
correction part of the calculation (scheme D). Horizontal lines represent sc GW results for comparison (do not depend on the basis set size for vertex). k-meshes $4\times4\times4$ have been used in both cases.} \label{bnd_conv}
\end{figure}

The convergence of the band width for Na and the band gap for Si with respect to the number of bands included in the vertex-related part of
calculations is shown in Fig.\ref{bnd_conv}. Here too one can see a striking difference between the convergence rate for the alkali metal on
the one hand and the semiconductor on the other. Whereas just 3-4 states closest to the chemical potential were enough to get the right band width in
Sodium, the convergence in Si happens only when one includes at least 30 band states (which still is almost 10 times smaller than the number of bands needed for GW part). It is important to mention, however, that not all properties
of Na show the same rate of convergence as the band width does. For example, the uniform polarizability, which was used to test how close the
calculated polarizability is to the physical one, was well converged only after inclusion of 15-20 bands in the case of Na.

As it can become clear from the formulae presented in the Appendix, the considerable (actually the most computationally expensive) part of the calculations is performed in the real space.
So it is important not only to take a certain number of bands into account for the vertex part, but also to represent them accurately with the smallest number of
orbitals (as compared to the full FLAPW+LO representation) inside muffin-tin spheres and with the smallest number of the real space mesh points in the interstitial region.

In case of Na and K the spd-basis was used in the MT-spheres for the vertex-related part of the calculations, i.e. 18 functions (both the solutions of radial equations $\varphi$ and their energy derivatives $\dot{\varphi}$ were always included in the basis set). In case of Si the sp-basis was used for both Si-atoms and empty-spheres (i.e. 32 functions in MT spheres altogether), which was good because Si structure is poorly packed and MT-spheres are small. In case of LiF the spd-basis was used for F, and the sp-basis for Li (26 functions totally). In all cases the uniform real-space mesh
$4\times4\times4$ in the unit cell was used to represent functions in the interstitial region. It was checked that the above parameters of the real-space
representation are good enough if one retains up to 25-30 bands in the vertex-related part of the calculation (with an estimated uncertainty 0.03$\div$0.05 eV in the calculated spectra). If one wants to increase the number
of bands included in the vertex part it would be necessary also to increase the accuracy of their real space representation. For comparison, the real space representation of the band states in the GW part of the calculations included orbitals up to $L_{max}=6$ inside MT spheres and the regular meshes $10\times10\times10$ ($12\times12\times12$ for Si) to represent the functions in the interstitial region.

\begin{figure}[b]
\centering
\includegraphics[width=8.5 cm]{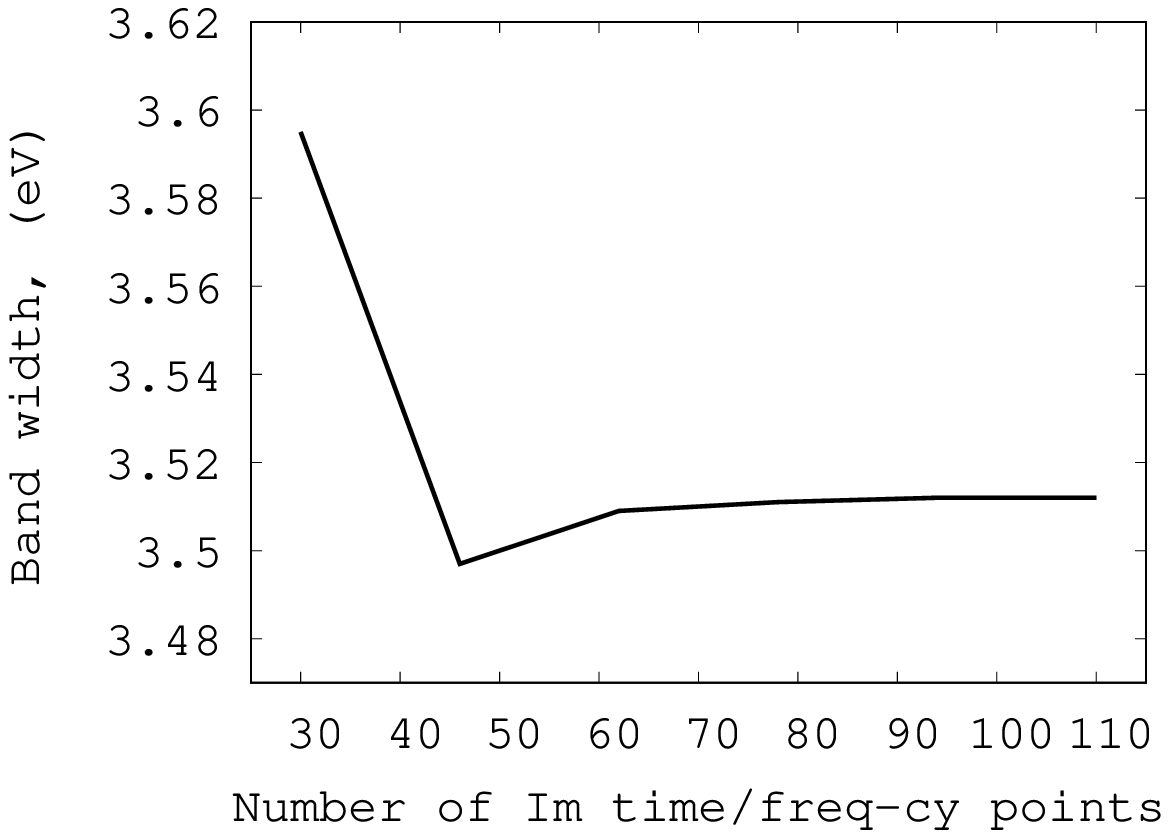}
\caption{Convergence of the band width (Na) with respect to the number of points on imaginary time/frequency mesh in scGW calculation. The temperature is 1000K. k-mesh is $12\times12\times12$.} \label{tau_na}
\end{figure}

\begin{figure}[t]
\centering
\includegraphics[width=8.5 cm]{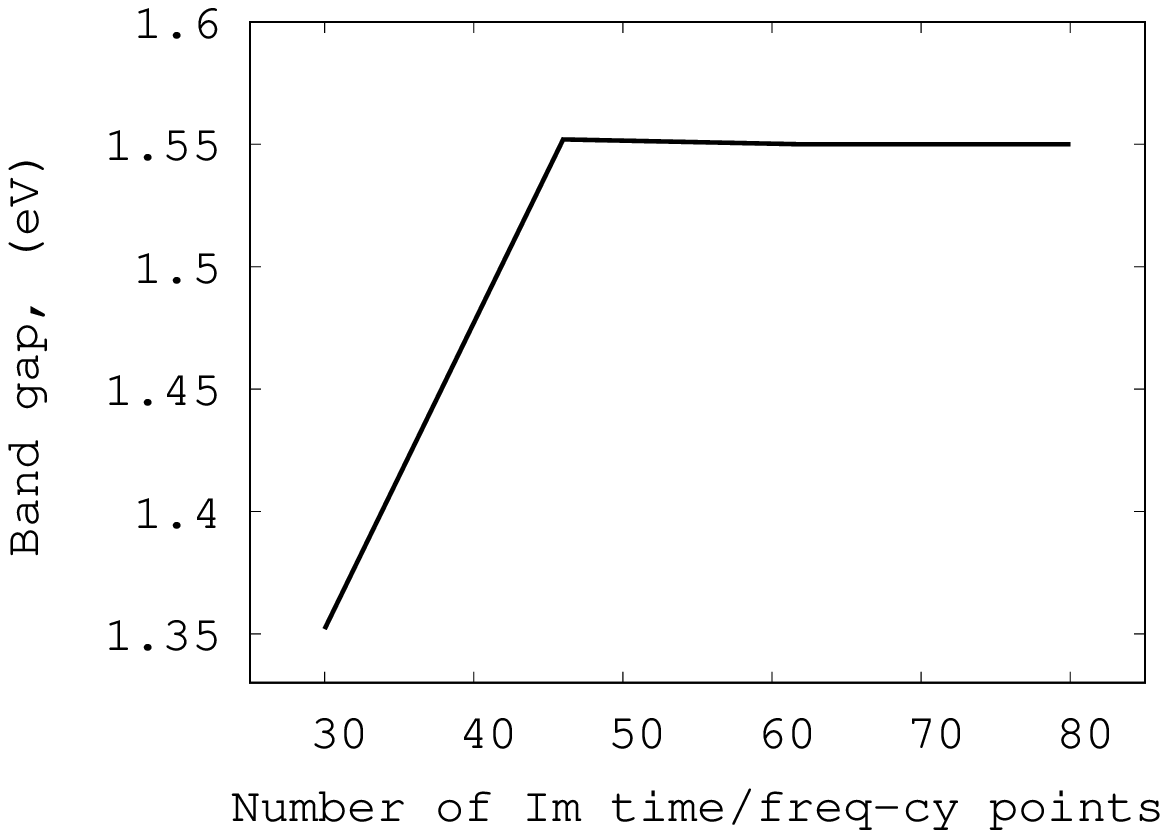}
\caption{Convergence of the band gap (Si) with respect to the number of points on imaginary time/frequency mesh in scGW calculation. The temperature is 1000K. k-mesh is $8\times8\times8$.} \label{tau_si}
\end{figure}

The convergence with respect to the number of imaginary time/frequency points is presented in Figs.\ref{tau_na} and \ref{tau_si}. The details about the meshes can be found in Ref.[\onlinecite{prb_85_155129}]. The number of imaginary time points and the number of frequency points was the same in the calculations, so only one variable is used in the figures. As it was already stated above, this number could be different in the GW part and in the vertex part. But this opportunity for optimization has not been implemented yet. As one can see, very good convergence is obtained beginning with approximately 60 points, which was used in all calculations presented in this work.

\begin{figure}[b]
\centering
\includegraphics[width=8.5 cm]{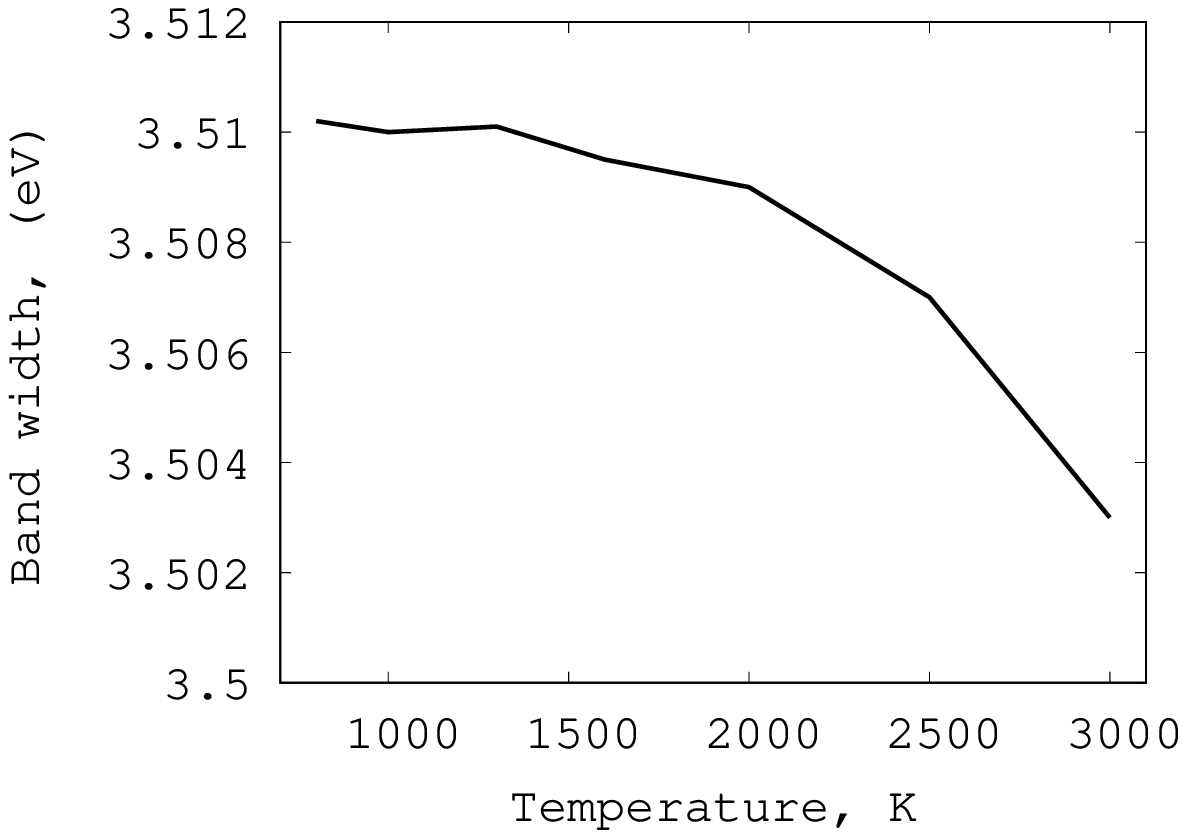}
\caption{Convergence of the band width (Na) with respect to the temperature in scGW calculation. k-mesh is $12\times12\times12$.} \label{temp_na}
\end{figure}

\begin{figure}[t]
\centering
\includegraphics[width=8.5 cm]{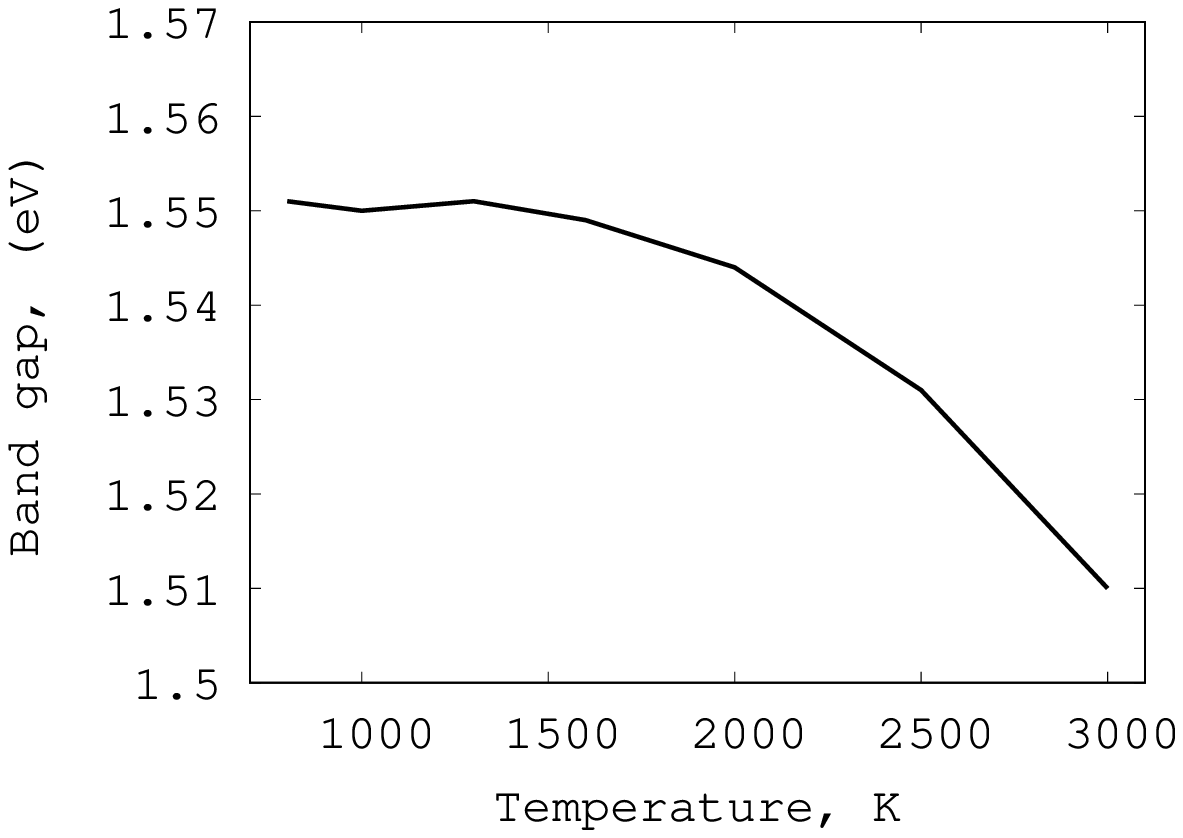}
\caption{Convergence of the band gap (Si) with respect to the temperature in scGW calculation. k-mesh is $8\times8\times8$.} \label{temp_si}
\end{figure}

Dependence of the results with respect to the electronic temperature is shown in Figs. \ref{temp_na} and \ref{temp_si}. As one can see, it is sufficiently weak for both materials. Corresponding uncertainty can be estimated to be not more than 0.002eV. In all presented below results the temperature was fixed at 1000K.

\begin{figure}[b]
\centering
\includegraphics[width=8.5 cm]{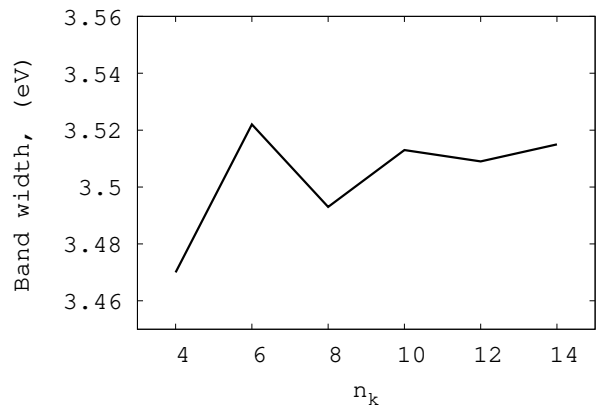}
\caption{Convergence of the band width (Na) with respect to $n_{k}=N_{k}^{1/3}$ with $N_{k}$ being the number of points in the Brillouin zone. The data are shown for scGW calculation.} \label{k_na}
\end{figure}

\begin{figure}[t]
\centering
\includegraphics[width=8.5 cm]{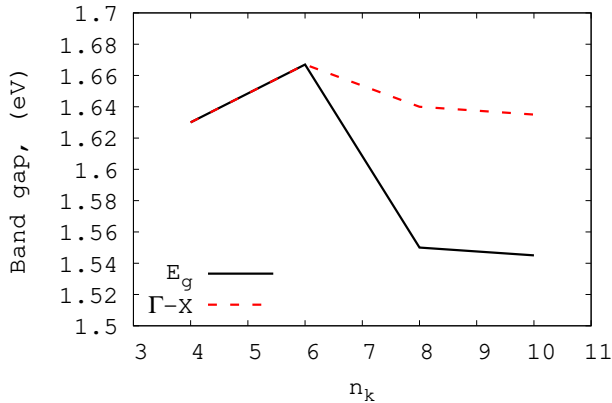}
\caption{(Color online) Convergence of the fundamental gap and band gap $\Gamma-X$ for Si with respect to $n_{k}=N_{k}^{1/3}$ with $N_{k}$ being the number of points in the Brillouin zone. The data are shown for scGW calculation.} \label{k_si}
\end{figure}

\begin{table*}[t]
\caption{Calculated one-electron energies at points of high symmetry for Si (in eV), together with available theoretical and experimental results. All theoretical results have been obtained in $G_{0}W_{0}$ approximation.} \label{g0w0_si}
\begin{center}
\begin{tabular}{@{}c c c c c c c c c c c c c}   & $\Gamma_{1v}$  & $\Gamma'_{25c}$ & $\Gamma_{15c}$  & $\Gamma'_{2c}$  & $X_{1v}$  & $X_{4v}$  & $X_{1c}$ & $L'_{2v}$ & $L_{1v}$ & $L'_{3v}$ & $L_{1c}$ & $L_{3c}$\\
\hline
 Ref.[\onlinecite{cpc_117_211}] &-11.57  &0.0  &3.24  &3.94  &-7.67  &-2.80  &1.34   &-9.39   &-6.86    &-1.17   &2.14    &4.05 \\
 Ref.[\onlinecite{prb_56_10228}] &-11.57  &0.0  &3.23  &3.96  &-7.57  &-2.83  &1.35   &-9.35   &-6.78    &-1.20   &2.18    &4.06 \\
 Ref.[\onlinecite{prb_67_155208}] &-11.85  &0.0  &3.09  &4.05  &-7.74  &-2.90  &1.01   &-9.57   &-6.97    &-1.16   &2.05    &3.83 \\
 Ref.[\onlinecite{prb_76_165106}] &-11.89  &0.0  &3.13  &4.02  &  &-2.96  &1.11   &   &    &-1.25   &2.05    &3.89 \\
 Ref.[\onlinecite{prb_87_155148}] &-11.64  &0.0  &3.25  &3.92  &-7.75  &-2.88  &1.36   &-9.38   &-6.93    &-1.23   &2.21    &4.00 \\
 Ref.[\onlinecite{prb_90_075125}] &-11.82  &0.0  &3.21  &  &  &-2.86  &1.22   &   &    &-1.21   &2.06   \\
Ref.[\onlinecite{prb_94_035118}] &  &0.0  &3.24  &  &  &-2.86  &1.25   &   &    & -1.22  & 2.09   & \\
This work  &-11.88  &0.0  &3.08  &3.96  &-7.73  &-2.93  &1.08  & -9.51  & -6.94   & -1.24  & 2.01 & 3.86  \\
Exp.[\onlinecite{cpc_117_211}]  &-12.5$\pm$0.6  &0.0  &3.40  &4.23  &  &-2.90  &1.25   & -9.3$\pm$0.4  & -6.7$\pm$0.2   & -1.2$\pm$0.2  & 2.1   & 4.15$\pm$0.1  \\
  &  &  &3.05  &4.1  &  &-3.3$\pm$0.2  &   &   &    &   & 2.4$\pm$0.1   &  
\end{tabular}
\end{center}
\end{table*}

One more opportunity to optimize the vertex part which has not been explored in this work is to use different meshes of points in the Brillouin zone for the GW part and for the vertex part.\cite{k_vrt_msh} All results presented in this work (if not specified) have been obtained
using $\mathbf{k}$-mesh $4\times4\times4$ in the Brillouin zone. Whereas this kind of mesh is not always good enough for GW part, it should be sufficient for the vertex part.\cite{k_vrt_msh} The convergence of the GW part has been checked by performing scGW calculations with larger number of $\mathbf{k}$-points (Figs. \ref{k_na} and \ref{k_si}). One point related to the band gap of Si should be clarified here. Fundamental gap in Si is measured between the highest occupied band at $\Gamma$ point in the Brillouin zone and the lowest unoccupied band at a certain point along $\Gamma-X$ line. However, when one uses coarse k-meshes (such as $4\times4\times4$ or $6\times6\times6$), it so happens that the lowest unoccupied band is exactly at $X$ point. It is easy to perform scGW calculations with sufficiently fine k-meshes and, thus, distinguish the fundamental gap and a gap between $\Gamma$ and $X$ points (from now on it will be called $\Gamma-X$ band gap). However, it is hard to take k-mesh finer than $4\times4\times4$ in vertex corrected calculations (at least presently). The values of both gaps are known from the experiment.\cite{prb_47_2130,jap_45_1846} So it is natural to compare the results from vertex-corrected calculations with experimental $\Gamma-X$  band gap, as it is done below in the table \ref{B_gap_si}. In the Fig. \ref{k_si}, however, both gaps are shown, and their difference converges to 0.09eV which is very close to the experimental difference 0.08eV. Having this said, one can now look at Figures \ref{k_na} and \ref{k_si} and estimate, that by using $4\times4\times4$ k-mesh one brings an uncertainty about 0.04eV in the band width of Na, and an uncertainty about 0.01eV in the calculated $\Gamma-X$ band gap of Si. Corresponding uncertainties for K and LiF were estimated to be 0.03eV and 0.05eV correspondingly. They are much smaller than the difference between the band widths/gaps obtained in the scGW and in vertex-corrected calculations and can be safely neglected in this study.

The analytic continuation of the correlation part of the self energy needed for the spectral function evaluation has been performed following
the scheme described before in the Appendix D of the Ref.[\onlinecite{prb_85_155129}]. The values of the small positive shift from the real
frequency axis were $2\div5\cdot10^{-3}$eV for the materials studied.

In vertex-corrected cases the scGW calculation (12-20 iterations till convergence) was performed before the vertex-related part of the calculation. In the
vertex part, six iterations in the small loop (Eqs. \ref{K0_def}-\ref{d_Vert3}) were sufficient to converge within $1\%$ in $\triangle\Gamma$ in
the cases of Si and LiF, which resulted in very good convergence of the band gaps. Slightly slower convergence was noticed in Na (8 iterations to
reach similar convergence) and in K (12 iterations). The number of iterations in the big loop (Eqs. \ref{Vert_0}-\ref{D4}) of the vertex part of
the calculation was 5-8 depending on the material, which provided good convergence of the band widths/gaps.

\begin{table}[b]
\caption{Average time per one iteration. 96 MPI processes were used.} \label{time}
\begin{center}
\begin{tabular}{@{}c c c c} Scheme   & Na/K  & Si & LiF  \\
\hline
A  & 16 seconds  & 530 seconds  & 33 seconds  \\
B  & 12 hours  & 3.5 hours  & 13 hours \\
E  & 20 hours  & 18 hours  & 50 hours
\end{tabular}
\end{center}
\end{table}

As a further test to check the performance of the code, the $G_{0}W_{0}$ (based on LDA) calculation of the electronic structure of Si has been performed. Results are shown in Table \ref{g0w0_si} where they are compared with earlier calculations and experiment. One shot ($G_{0}W_{0}$) includes all ingredients of scGW calculation and, thus, is useful to check the implementation of GW part. The difference between one-electron energies from present work and earlier calculations is, generally, very small, testifying the adequacy of numerical approximations made in this study.

It is interesting how the computer time increases when one includes vertex corrections of different complexity. Table \ref{time} provides the time per one iteration. k-mesh $4\times4\times4$ has been used, so GW shows a good performance. As one can see, inclusion of higher order diagrams makes calculations a lot more time consuming. However, the vertex part of the code has not yet been totally optimized. With the optimizations mentioned earlier and other improvements in the code the times should be reduced by the factor of 10 or more.

\section{Results}
\label{res}

In this section the results from self consistent calculations are presented. They are compared with earlier self consistent calculations and with experimental data. In order to make comparison with earlier calculations more meaningful, present fully self-consistent calculations have been supplemented with QSGW and QSGW$_{0}$ calculations using the same computer code. Partially self-consistented GW have also been included ($GW_{LDA}$ in the Table \ref{Bw_na_k} below). In case of Na and K, G$_{0}$W$_{0}$ calculations have also been performed for comparison with previous works. Whenever G$_{0}$ was needed it was evaluated within LDA with parametrization from Ref.[\onlinecite{prb_45_13244}]. The details of the implementation of quasi-particle self consistence on imaginary axis have been described before.\cite{prb_85_155129}

In metallic cases (Na and K) the calculations based on the scheme F appeared to be unstable (because of the above mentioned inconsistency between the kernel of the Bethe-Salpeter equation and diagrammatic representation of the self energy). So the corresponding results are missing below. For the insulating materials scheme F seems to be acceptable, which, however, might be just because the higher order diagrams in the self energy are less essential for Si and LiF.

\subsection{Na and K} \label{Na_K}

\begin{table}[t]
\caption{Band widths of Na and K (eV).} \label{Bw_na_k}
\begin{center}
\begin{tabular}{@{}c c c c} Method   & Ref.  & Na & K  \\
\hline\hline
G$_{0}$W$_{0}$  & [\onlinecite{prb_86_035120}]  &2.887 & \\
            & [\onlinecite{prb_89_081108}]  &3.00 & \\
            & This work  &3.02 &1.90 \\
            \hline
QSGW        & [\onlinecite{prl_96_226402}]  &3.0 & \\
            & This work  &3.17 &1.95 \\
            \hline
QSGW$_{0}$  & This work  &2.87 &1.72\\
            \hline
GW$_{LDA}$        & [\onlinecite{prl_59_819}]  &2.5 & \\
            & [\onlinecite{prb_38_5976}]  & &1.58$\pm$0.1 \\ 
            & [\onlinecite{prb_89_081108}]  &2.83 & \\
        & [\onlinecite{prb_86_035120}]  &2.673 & \\
GW$\Gamma_{LDA}$        & [\onlinecite{prb_86_035120}]  &2.958 & \\
            \hline
A  & This work  &3.47  &2.38  \\
B  & This work  &3.03  &2.04 \\
C  & This work  &3.24  &2.16  \\
D  & This work  &2.73  &1.69  \\
E  & This work  &2.71  &1.71  \\
G  & This work  &2.82  &1.84  \\
            \hline
Exp.  &[\onlinecite{prl_60_1558,prb_41_8075}]  & 2.65  & 1.60$\pm$0.05
\end{tabular}
\end{center}
\end{table}

\begin{figure}[b]
\centering
\includegraphics[width=8.5 cm]{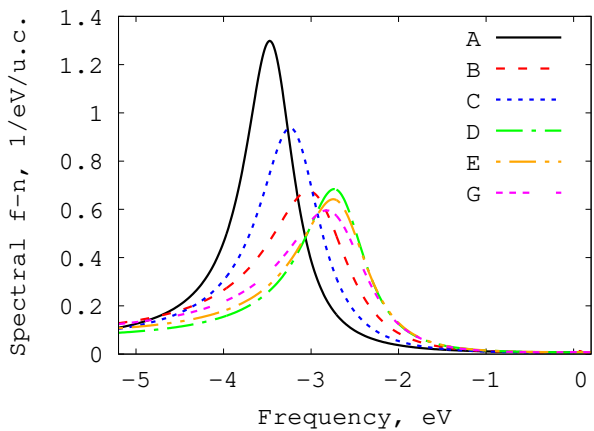}
\caption{(Color online) Spectral function of Na at $\Gamma$ point in the Brillouin zone. Chemical potantial corresponds to zero frequency.} \label{a_G_na}
\end{figure}

Before presenting results of fully sc calculations (without and with vertex corrections) let us look at the results for Na and K obtained with simplified GW schemes. The results from G$_{0}$W$_{0}$ and QSGW calculations are included in Table \ref{Bw_na_k} where they are compared to the similar calculations performed with the code being presented in this work. The band widths obtained in different G$_{0}$W$_{0}$ calculations are pretty close to each other and are in reasonable agreement with experiment. 

QSGW study by Schilfgaarde et al\cite{prl_96_226402} for Na shows $15\%$ too wide band width, and similar calculation of this work gives even slightly larger deviation from experiment. Two results are slightly different from each other, which most likely is because of the linearization of self energy in QSGW approach of the present work. The best results among simplified GW schemes without vertex corrections provides QSGW$_{0}$ method. However, similar to the G$_{0}$W$_{0}$ approach, it depends on the starting point which makes its predictive power questionable.

Band widths obtained with simplified two-point LDA vertex\cite{prl_59_819,prb_38_5976,prb_86_035120,prb_89_081108} are also included in Table \ref{Bw_na_k}, where GW$_{LDA}$ means that LDA vertex included only in W and GW$\Gamma_{LDA}$ includes LDA vertex also in self energy. Good agreement with experimental band width is obtained in GW$_{LDA}$ approach, whereas the inclusion of vertex correction in self energy deteriorates the results.

Opposite to the studies based on the quasi-particle sc, where band width can easily be found by looking at the corresponding quasi-particle energies, in fully sc approaches one has to analyze corresponding spectral functions.
The band width of alkali metals is defined by the position of the valence band bottom at the $\Gamma$-point in the Brillouin zone relative to the
chemical potential. So, in this work it was found from the position of the peak of the spectral function corresponding to the
$\Gamma$-point. As an example, the spectral function of Sodium is shown in Figure \ref{a_G_na}. Let us now look at the results of fully sc calculations also presented in Table \ref{Bw_na_k}. As one can see, for both metals the vertex corrected schemes (D, E, and G) provide $5\div10$ times better
accuracy than scGW approach. The schemes B (vertex $\Gamma_{1}$ in both $P$ and $\Sigma$) and C (vertex from Bethe-Salpeter equation in $P$, but
no vertex correction in $\Sigma$) show worse performance and correct only $30-50\%$ of the scGW error. The small remaining error in D, E, and G schemes
most likely could be reduced farther if the basis set was better in the vertex-related part of the calculations (i.e. if the representation of the
bands in the real space could be better). For example the band width of Na obtained with only the sp-basis inside MT spheres
(vertex-related part of the calculation) was $2.85$eV in the scheme D, i.e. the extension to the spd-basis resulted in $0.12$eV improvement. Higher order diagrams not included in this study can also be a reason for the remaining errors. As it can be seen from the Table \ref{Bw_na_k}, schemes D, E, and G are superior in accuracy if one compares them with QSGW approximation.

Potassium is the next (after Sodium) alkali metal in the Periodic table and, naturally, the calculations show similar tendencies in its properties. However, K is slightly more correlated than Na, as one can understand drawing the parallel between these two metals and the electron gas with two corresponding densities. Valence electron density in Potassium is lower than in Sodium,
and the electron gas with lower density is more correlated. It is also seen from the row A in the Table \ref{Bw_na_k}: in the case of Na the error of scGW
approach is $\sim$30\% whereas it is $\sim$48\% in the case of K. Stronger correlations in K can also be seen from the comparison of the rows E and G in the same Table: the neglect of spin-flip diagrams has larger effect in K than in Na. Also, iterative solution of the Bethe-Salpeter equation converges slower in case of K. Nevertheless, vertex corrected schemes D and E allow to reach good accuracy in the calculated band width of Potassium as well as of Sodium.

Thus, one can conclude that in both alkali metals it is imperative to include vertex corrections both in the polarizability (Bethe-Salpeter equation
has to be solved) and in the self energy (with the first order vertex). However, an additional care should be taken if one wants to include higher order vertex corrections in the self energy: the kernel in the Bethe-Salpeter equation should also be modified in this case.

\begin{figure}[t]
\centering
\includegraphics[width=8.5 cm]{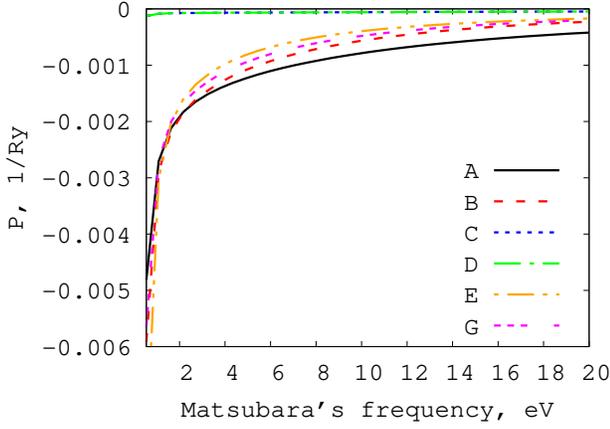}
\caption{(Color online) Uniform polarizability ($P^{\mathbf{q}=0}_{\mathbf{G}=\mathbf{G}'=0}(\nu)$) of Na as a function of Matsubara's frequency.} \label{P_0_na}
\end{figure}

\begin{figure}[b]
\centering
\includegraphics[width=8.5 cm]{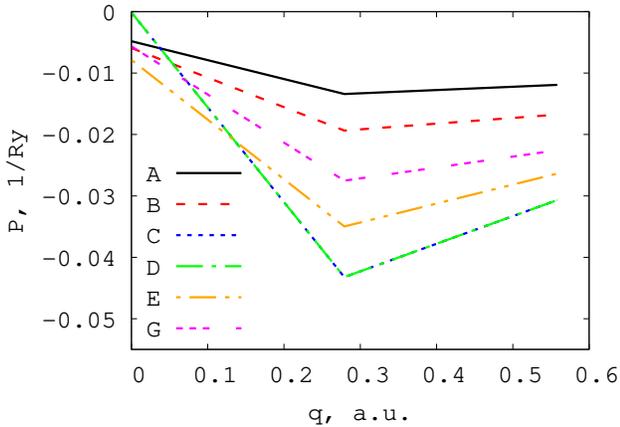}
\caption{(Color online) Polarizability ($P^{\mathbf{q}}_{\mathbf{G}=\mathbf{G}'=0}(\nu=\frac{2\pi}{\beta})$) of Na for the smallest positive Matsubara's frequency as a function of $q=|\mathbf{q}|$ along the $\Gamma$-$N$ line in the Brillouin zone. $\beta$ stands for the inverse temperature.} \label{P_q_na}
\end{figure}

\begin{figure}[t]
\centering
\includegraphics[width=9.0 cm]{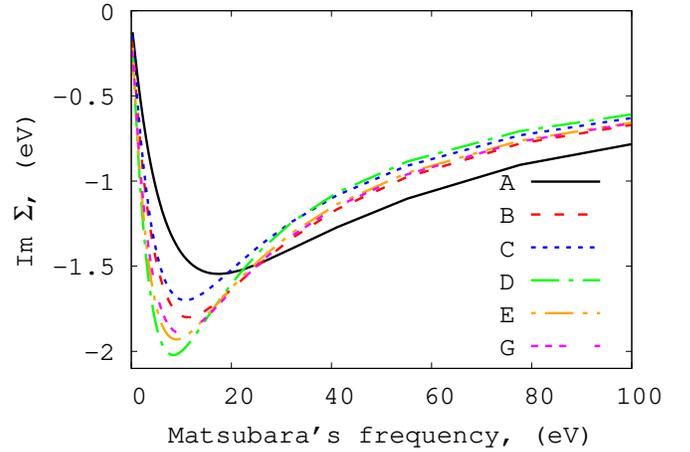}
\caption{(Color online) Imaginary part of self energy of Na at $k=(0;0;0)$ as a function of Matsubara's frequency.} \label{sig_G_Im_na}
\end{figure}

A few technical details are presented below, which can be useful for future development of the method. They are quite similar for all four materials, so Na is used as an example.

Figure \ref{P_0_na} shows the homogeneous ($P^{\mathbf{q}=0}_{\mathbf{G}=\mathbf{G}'=0}(\nu)$) component of the polarizability of Na as a function of
positive Matsubara's frequencies $\nu$. If the polarizability is exact or if it is not exact but "physical", this function should be zero for all
$\nu\neq0$ (for metals). There are two approaches (C and D) in this study where the polarizability is physical (it is actually the same in C and D by
construction). Thus Fig.\ref{P_0_na} provides an indication that numerical approximations are good enough making the lines C and D almost
identically zero. First order conserving scheme B shows steady improvement as compared to the scGW for all frequencies but the first two, where it
is even slightly worse than scGW result. Similar behavior shows the scheme E, which is only slightly better than scheme B at intermediate
frequencies. But considerable improvement in the spectral function obtained with the scheme E compared to the spectral function in the scheme B tells us that the long wave limit of
the polarizability is not very important for the one-electron spectral properties. More important is the behavior of the polarizability in the whole Brillouin zone, as it follows from the next paragraph.

\begin{figure}[b]
\centering
\includegraphics[width=8.5 cm]{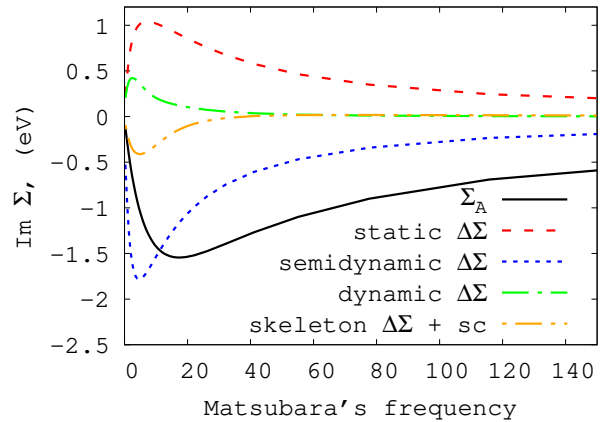}
\caption{(Color online) Components of the self energy (scheme D, bottom of the valence band) for Na. 'Skeleton $\triangle\Sigma$ + sc' is obtained as a sum $\triangle\Sigma^{static}+\triangle\Sigma^{semidynamic}+\triangle\Sigma^{dynamic}$ (see text for the details). $\Sigma_{A}$ stands for the self energy in the scheme A.} \label{s_comp_na}
\end{figure}

In the Figure \ref{P_q_na} the polarizability $P^{\mathbf{q}}_{\mathbf{G}=\mathbf{G}'=0}(\nu=2\pi/\beta)$ for the smallest positive Matsubara's frequency is
presented as a function of $|\mathbf{q}|$ along the $\Gamma$-$N$ line in the Brillouin zone. As one can see, there is a certain correlation between the average amplitude
of the polarizability in the Brillouin zone and the band width. Namely, among the schemes with similar diagrammatic representation of the self energy (B, D, E, and G) the tendency in the average amplitude of the polarizability ($P_{B}<P_{G}<P_{E}<P_{D}$) follows the opposite tendency in the calculated band width error $\varepsilon$ ($\varepsilon_{B}>\varepsilon_{G}>\varepsilon_{E}\approx\varepsilon_{D}$). If, however, one compares the band widths in the
schemes C and D (which have identical polarizabilities but scheme C doesn't include vertex correction to the self energy) one will realize the importance of the vertex correction in the self energy.

Imaginary part of the self energy  at the $\Gamma$ point in the Brillouin zone (diagonal matrix element corresponding to the bottom of the valence band)
is presented in the Figure \ref{sig_G_Im_na}. Self energy includes the vertex corrections indirectly (through $W$) and directly through the
skeleton diagrams in the self energy itself. As a result it correlates with the final band width stronger than the polarizability. As one can see from the
figure and from the Table \ref{Bw_na_k}, the larger amplitude of the self energy corresponds to the smaller band width and vice versa.

Figure \ref{s_comp_na} presents different components of the imaginary part of $\Sigma$ obtained in the scheme D. Corresponding skeleton diagram can be written schematically as $\triangle\Sigma=WWGGG$. As it is explained in more details in the Appendix \ref{Sig_corr}, the separation of $W$ into bare Coulomb (V) and screening $\widetilde{W}$ interactions ($W=V+\widetilde{W}$) results in three components of $\triangle\Sigma$: static ($\triangle\Sigma^{static}=VVGGG$), semidynamic ($\triangle\Sigma^{semidynamic}=[\widetilde{W}V+V\widetilde{W}]GGG$), and dynamic ($\triangle\Sigma^{dynamic}=\widetilde{W}\widetilde{W}GGG$). The line marked as 'Skeleton $\triangle\Sigma$ $+$ sc' in Fig.\ref{s_comp_na} represents the sum of these three  contributions. The addition '$+$ sc' means that the skeleton $\triangle\Sigma$ diagram has been evaluated with fully sc G and W. $\Sigma_{A}$ line (GW diagram in the scheme A) is given for comparison.

First of all one has to stress the importance of full dynamical treatment of W (frequency dependence). As it is seen, the individual components of the skeleton $\triangle\Sigma$ diagram are of the same magnitude as $\Sigma_{A}$. However, their sum is much smaller (about 4 times smaller than $\Sigma_{A}$) and is very localized in frequency space (it is almost negligible for $\omega>20$eV whereas $\Sigma_{A}$ is pretty large up to a few hundred of eV's). A few calculations have been performed with only the static $\triangle\Sigma$ included, which was evaluated using the static interaction equal to i)the bare Coulomb V, ii)$W(\nu=0)$, and iii)a few $W$'s at intermediate $\nu$'s. $Im \triangle\Sigma$ in the calculations with reduced static interaction (as compared to the V) was qualitatively similar as $\triangle\Sigma^{static}$ presented in the figure, but with reduced amplitude. All curves were positive, whereas the right one (shown as 'Skeleton $\triangle\Sigma$ $+$ sc' in the Figure) obtained with proper dynamic $W$ is negative. Corresponding effect on the band width was also positive: all calculations with static W's resulted in increased band width, whereas the proper treatment of frequency dependence in $W$ results in the reduced band width. Similar findings were discovered in other materials studied in this work. This essentially explains why the authors of Ref.[\onlinecite{prl_112_096401}] were obtaining the increase in band gaps when they applied vertex correction to the self energy evaluated with static $W$.

\subsection{Si and LiF} \label{Si_LiF}

\begin{figure}[t]
\centering
\includegraphics[width=8.5 cm]{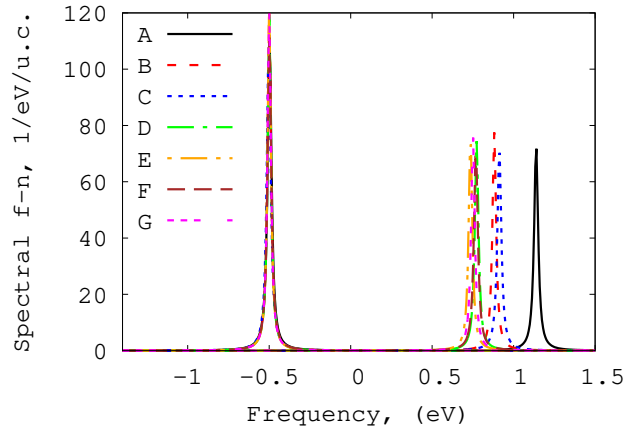}
\caption{(Color online) Spectral function of Si at $\Gamma$ (lines below zero) and $X$ (lines above zero) points in the Brillouin zone. For
convenience, all lines have been shifted to place the highest occupied state energy at -0.5eV for all approaches.} \label{a_g_x_si}
\end{figure}

\begin{figure}[b]
\centering
\includegraphics[width=8.5 cm]{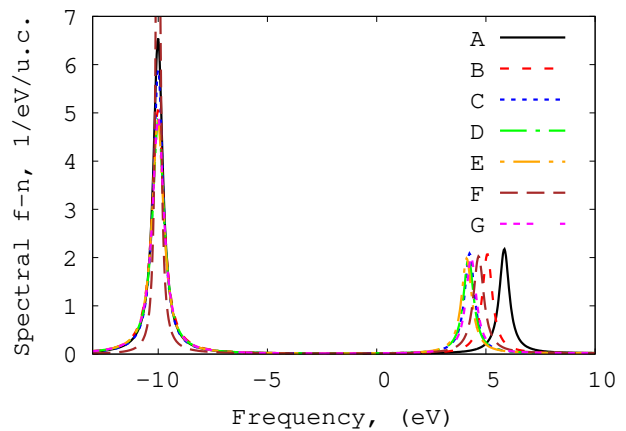}
\caption{(Color online) Spectral function of LiF at the $\Gamma$ point in the Brillouin zone.} \label{a_g_lif}
\end{figure}

\begin{table}[b]
\caption{$\Gamma$-X band gap and fundamental gap (E$_{g}$) of Si (eV). Screened interaction W was fixed at RPA level (calculated with G from LDA or PBE) in QSGW$_{0}$ approach. PBE stands for Perdew-Burke-Ernzerhof functional\cite{prl_77_3865}. Vertex corrections were included in W (through effective kernel $f_{xc}$) in QSGW$_{e-h}$. Calculations with the scheme A have been performed for k-mesh $8\times8\times8$ to show the difference between $\Gamma$-X band gap and fundamental gap in scGW method.} \label{B_gap_si}
\begin{center}
\begin{tabular}{@{}c c c c} Method   & Ref.  & $\Gamma$-X gap & E$_{g}$ \\
\hline
            \hline
QSGW        & [\onlinecite{prb_76_165106}]  &1.37  & 1.23 \\
            & [\onlinecite{prl_99_246403}]  &  & 1.41 \\
            & [\onlinecite{prb_92_041115}]  &  & 1.47 \\
            & This work  &1.50  & 1.41 \\
            \hline
QSGW$_{0}$,(PBE)  & [\onlinecite{prl_99_246403}]  &  & 1.28 \\
       (LDA)     & [\onlinecite{prb_90_075125}]  &1.22&  \\
       (PBE)     & [\onlinecite{prb_90_075125}]  &1.28&  \\
      (PBE)      & [\onlinecite{prb_92_041115}]  &  & 1.28 \\
      (PBE)      & [\onlinecite{prb_93_115203}]  &  & 1.19 \\
    (LDA)        & This work  &1.24  &1.15 \\
            \hline
QSGW$_{e-h}$  & [\onlinecite{prl_99_246403}]  &  & 1.24 \\
              & [\onlinecite{prb_92_041115}]  &  & 1.30 \\
            \hline
A  & This work  &1.63  &  \\
A,$8\times8\times8$  & This work  &1.64  &1.55  \\
B  & This work  &1.38  &  \\
C  & This work  &1.41  &  \\
D  & This work  &1.27  &  \\
E  & This work  &1.24  &  \\
F  & This work  &1.27  &  \\
G  & This work  &1.25  &  \\
            \hline
Exp.  &[\onlinecite{prb_47_2130,jap_45_1846}]  &1.25  &1.17 
\end{tabular}
\end{center}
\end{table}

\begin{table}[t]
\caption{Band gap of LiF (eV). Screened interaction W was fixed at RPA level (calculated with G from LDA or PBE\cite{prl_77_3865}) in QSGW$_{0}$ approach.} \label{B_gap_lif}
\begin{center}
\begin{tabular}{@{}c c c} Method   & Ref.  & Band gap  \\
\hline\hline
QSGW        & [\onlinecite{prb_75_235102}]  &15.10  \\
            & [\onlinecite{prb_90_165133}]  &16.17  \\
            & This work  & 16.57 \\
            \hline
QSGW$_{0}$, (PBE)  & [\onlinecite{prb_75_235102}]  &13.96 \\
    (PBE)        & [\onlinecite{prb_90_165133}]  &14.29 \\
            & [\onlinecite{prb_89_125429}]  &13.62   \\
    (LDA)        & This work  & 14.76 \\
            \hline
A  & This work  &15.85    \\
B  & This work  &15.06    \\
C  & This work  &14.25    \\
D  & This work  &14.21    \\
E  & This work  &14.12    \\
F  & This work  &14.56    \\
G  & This work  &14.32    \\
\hline
Exp.  &[\onlinecite{prb_13_5530}]  & 14.2$\pm$0.2
\end{tabular}
\end{center}
\end{table}

Spectral functions of Si and LiF are presented in Figures \ref{a_g_x_si} and \ref{a_g_lif} correspondingly. As it was explained before, for the k-mesh $4\times4\times4$ the band gap in Silicon is measured between the highest occupied state at the $\Gamma$-point in the Brillouin zone and the lowest unoccupied state at the $X$-point. Correspondingly, the spectral function at these two $\mathbf{k}$-points have been combined in the figure \ref{a_g_x_si}. The band gap in LiF corresponds to the direct transition between the highest occupied and the lowest unoccupied band at the $\Gamma$-point in the
Brillouin zone. Tables \ref{B_gap_si} and \ref{B_gap_lif} present numerical data for the band gaps in Si and LiF compared with the experiment and with earlier calculations. First of all, one can check that for QSGW and QSGW$_{0}$ methods the band gaps of Si obtained in this study are within the range of results obtained in earlier studies. This confirms that numerical accuracy of the GW part of the code is sufficiently good. Slight increase in the band gap of LiF relative to the earlier results most likely is attributed to the fact that quasiparticle approach in this work involves linearization of self energy.\cite{prb_85_155129} This linearization is not a part of the fully sc methods (A-G), studied in this work. Comparison with experimental data shows that, similar to alkali metals, QSGW$_{0}$ is superior in accuracy among quasiparticle-based sc schemes.

Considering fully sc approximations, one can conclude from Tables \ref{B_gap_si} and \ref{B_gap_lif} that fully scGW is not very successful approach (the deviation from experiment in this case is more than 30\% for Si and more than 10\% for LiF). However, this error can be reduced practically to zero (within uncertainty of experiment) in both cases if one applies appropriate vertex corrections (schemes D, E, and G). It is interesting, that in case of LiF (wide gap insulator) the vertex corrections to the self energy are not very important (scheme C results in essentially the same band gap as the schemes D, E, and G). At the same time, the first order vertex correction in Polarizability (scheme B) is not sufficient: it is essential to solve Bethe-Salpeter equation for the polarizability. As for the scheme F which includes higher order diagrams in the self energy, the calculation was stable (as compared to alkali metals) but the band gap obtained shows worse accuracy for LiF than the results from D, E, and G schemes.

\section*{Conclusions}
\label{concl}

In conclusion, a few self-consistent schemes of solving the Hedin equations have been introduced. The combination of features which
distinguishes these schemes from the previously published works on the subject is the following: they are diagrammatic and self-consistent, they do not apply the
quasi-particle approximation for the Green function, they threat full frequency dependence of the interaction W.

For the materials studied in this work (Na, K, Si, and LiF) one can conclude that the vertex corrections both in the
polarizability and in the self energy are important. However, the vertex function which should be used in P has to be found from the Bethe-Salpeter equation,
whereas it is enough for the vertex function to be of the first order (in W) to make proper corrections in $\Sigma$. Inclusion of higher order diagrams in the self energy has to be supplemented with the corresponding increase in the complexity of kernel of the Bethe-Salpeter equation. Otherways their inclusion can make the whole scheme unstable which was the case for Na and K in this study.

The importance of proper treatment of the frequency dependence of $W$ in the vertex correction diagrams for the self energy has been revealed. It explained the increase in the calculated band gaps obtained in earlier works where static $W$ was used to evaluate the second order diagrams for the self energy.

The best schemes in this work allow to considerably improve the accuracy of the calculated band widths and band gaps: the error becomes 10 times
(or more) smaller than in the self-consistent GW approximation. Moreover, they show superior accuracy as compared to the commonly used nowadays QSGW approximation.

From the computational point of view, a few possible technical optimizations have been pointed out (different k/time/frequency-meshes for GW and vertex parts). In addition, one can take an advantage of the fact that the scheme D is one of the best in this study, and, as compared to another successful scheme (E), is far more efficient, because Bethe-Salpeter's equation should be solved only once in the scheme D, whereas in the scheme E it should be solved on every iteration. Another simplification, which worked sufficiently well in this study for Si and LiF, is to neglect spin-flips diagrams in the kernel of the Bethe-Salpeter equation. It also saves computation time considerably.

\section*{Acknowledgments}
\label{acknow}

This work was   supported by the U.S. Department of energy, Office of Science, Basic
Energy Sciences as apart of the Computational Materials Science Program. The calculations have been performed on the Dell HPC Cluster ELF I at the Rutgers University. I thank Koji Tanaka for the help with adapting the code for the ELF I. Fruitful discussions of the parallelization strategies with Viktor Oudovenko are highly appreciated.

\appendix

\section{Details of the vertex corrections evaluation}\label{vrt_det}

In this Appendix the details of the formulae are given in a form close to the implementation in the code. One notion should be mentioned here before proceeding. The functions ($K^{0}, \triangle K, \triangle \Gamma, Q, T$) which are evaluated in the course of iterations (\ref{K0_def}-\ref{d_Vert3}) are three-point functions. One of the three points can be considered as independent. In the representation accepted in this work, the independent point corresponds to the indexes $s, \mathbf{q}, \nu$, which are the reduced product basis index, the point in the Brillouin zone, and Matsubara's frequency (see below for the specifications). The calculations for every triplet of these indexes are totally independent, which is used to perform the calculations in parallel. Besides, one needs to do the calculations only for the irreducible set of $\mathbf{q}$-points. Having the iterations (\ref{K0_def}-\ref{d_Vert3}) converged, one can proceed with the corrections to the polarizability and to the self energy. For the evaluation of the latter, however, one needs to combine the information from the above triplets of indexes.

\subsection{Notations} \label{not}

In order to make the reading of the following sections easier, the notations have been collected here:

\begin{itemize}
    \item $\alpha$ - spin index
    \item $\lambda, \lambda', \lambda'', \lambda'''$ - band indexes. Bands obtained in the effective Hartree-Fock problem\cite{prb_85_155129} are used in the vertex part. See the section I in Ref.[\onlinecite{prb_85_155129}] for the details.
    \item $\mathbf{k}, \mathbf{q}$ - points in the Brillouin zone
    \item $s, s', s'', s_{1}, s_{2}$ - reduced product basis (RPB) index. When it is used together with vector $\mathbf{q}$ in the Brillouin zone
     (corresponding RPB function is $\Pi^{\mathbf{q}}_{s}$), it runs
    over all RPB (muffin-tins plus interstitial). When it is used together with atomic index $\mathbf{t}$
    (corresponding RPB function in this case is $\Pi^{\mathbf{t}}_{s}$) it runs over the part of full RPB belonging to the given atom.
    \item $\omega, \omega'$ - fermionic Matsubara's frequency
    \item $\nu$ - bosonic Matsubara's frequency
    \item $\tau, \tau'$ - Matsubara's time
    \item $\mu$ - chemical potential
    \item $\epsilon^{\alpha\mathbf{k}}_{\lambda}$ - band energies
    \item $\Psi^{\alpha\mathbf{k}}_{\lambda}$ - band wave functions
    \item $\beta$ - inverse temperature
    \item $\mathbf{R}$ - vectors of translations in real space
    \item $\mathbf{t}, \mathbf{t}'$ - coordinates (or indexes) of atoms in unit cell
    \item $L, L', L'', L'''$ - indexes combining orbital moment $l$, its projection $m$, and other quantum numbers distinguishing the orbitals
    $\phi^{\alpha}_{\mathbf{t}L}$ for given spin $\alpha$ and atom $\mathbf{t}$ ($L$-indexes also distinguish between $\phi$ and $\dot{\phi}$)
    \item $N_{\mathbf{k}}$ - full number of $\mathbf{k}$-points in the Brillouin zone
    \item $\omega_{\mathbf{q}}$ - geometrical weight of the $\mathbf{q}$-point in the Brillouin zone, i.e. the ratio of the number of vectors in
    the star of $\mathbf{q}$ and the full number of points in the Brillouin zone
    \item $\mathbf{r}, \mathbf{r}'$ - the points on the regular real space mesh in the unit cell
    \item $\mathbf{G}$ - reciprocal lattice vectors
    \item $\mathbf{G}_{s}$ - reciprocal lattice vector associated with reduced product basis index $s$.
\end{itemize}

\subsection{ $K^{0}$ calculation} \label{K0_calc}

Expanding $G$ in (\ref{K0_def}) in the band states, one gets the formulae

\begin{align}\label{K0_1}
&K^{0\alpha\mathbf{k}}_{\lambda\lambda'}(s\mathbf{q};\omega;\nu)=\nonumber\\&
-\sum_{\lambda''\lambda'''}G^{\alpha\mathbf{k}}_{\lambda\lambda''}(\omega)
\langle\Psi^{\alpha\mathbf{k}}_{\lambda''}|\Psi^{\alpha\mathbf{k}-\mathbf{q}}_{\lambda'''}\Pi^{\mathbf{q}}_{s}\rangle
G^{\alpha,\mathbf{k}-\mathbf{q}}_{\lambda'''\lambda'}(\omega-\nu),
\end{align}
and
\begin{align}\label{K0_2}
&K^{0\alpha\mathbf{k}}_{\lambda\lambda'}(s\mathbf{q};-\omega+\nu;\nu)=\nonumber\\&
-\sum_{\lambda''\lambda'''}G^{^{*}\alpha\mathbf{k}}_{\lambda''\lambda}(\omega-\nu)
\langle\Psi^{\alpha\mathbf{k}}_{\lambda''}|\Psi^{\alpha\mathbf{k}-\mathbf{q}}_{\lambda'''}\Pi^{\mathbf{q}}_{s}\rangle
G^{^{*}\alpha,\mathbf{k}-\mathbf{q}}_{\lambda'\lambda'''}(\omega),
\end{align}
with $\Pi^{\mathbf{q}}_{s}$ representing the product basis functions defined on the reduced set of band states. As it will be clear from the equations below, one needs to evaluate (\ref{K0_1}) and (\ref{K0_2}) for $\omega\geqslant\nu/2, \nu>0$ only. Two functions are needed to handle strong oscillations in $\tau$-dependence of $K(\tau,\nu)$ (see Eqn. (\ref{K0_6}) below).

Equation (\ref{d_Vert3}) is convenient to evaluate in real space and ($\tau;\nu$)-representation. Before transforming $K$ to the
$(\tau;\nu)$-representation, the Hartree-Fock contributions are subtracted

\begin{align}\label{K0_3}
K^{0,HF,\alpha\mathbf{k}}_{\lambda\lambda'}&(s\mathbf{q};\omega;\nu)=\nonumber\\ -&\frac{
\langle\Psi^{\alpha\mathbf{k}}_{\lambda}|\Psi^{\alpha\mathbf{k}-\mathbf{q}}_{\lambda'}\Pi^{\mathbf{q}}_{s}\rangle}
{(i\omega+\mu-\epsilon^{\alpha\mathbf{k}}_{\lambda})(i(\omega-\nu)+\mu-\epsilon^{\alpha\mathbf{k}-\mathbf{q}}_{\lambda'})},
\end{align}
and

\begin{align}\label{K0_4}
&K^{0,HF,\alpha\mathbf{k}}_{\lambda\lambda'}(s\mathbf{q};-\omega+\nu;\nu)=\nonumber\\ -&\frac{
\langle\Psi^{\alpha\mathbf{k}}_{\lambda}|\Psi^{\alpha\mathbf{k}-\mathbf{q}}_{\lambda'}\Pi^{\mathbf{q}}_{s}\rangle}
{(-i(\omega-\nu)+\mu-\epsilon^{\alpha\mathbf{k}}_{\lambda})(-i\omega+\mu-\epsilon^{\alpha\mathbf{k}-\mathbf{q}}_{\lambda'})}.
\end{align}

After subtraction one uses (\ref{tt_1}):

\begin{align}\label{K0_5}
K^{0\alpha\mathbf{k}}_{\lambda\lambda'}(s\mathbf{q};\tau;\nu)&= \frac{1}{\beta}\sum_{\omega\leqslant\nu/2}e^{-i\omega\tau}
\underbrace{K^{0\alpha\mathbf{k}}_{\lambda\lambda'}(s\mathbf{q};\omega;\nu)}_{large~at~\omega=0}
\nonumber\\&+\frac{1}{\beta}\sum_{\omega\geqslant\nu/2}e^{-i\omega\tau}
\underbrace{K^{0\alpha\mathbf{k}}_{\lambda\lambda'}(s\mathbf{q};\omega;\nu)}_{large~at~\omega=\nu}
\end{align}

In the first term strong oscillations in $K$ as a function of $\tau$ originating from $\omega\sim0$ are damped by exponential factor which has
weak $\tau$-dependence near $\omega=0$. In the second term the oscillations come from $\omega\sim\nu$, so one has to ensure the damping by rearranging
the exponential factors as the following

\begin{align}\label{K0_6}
K^{0\alpha\mathbf{k}}_{\lambda\lambda'}&(s\mathbf{q};\tau;\nu)=\underbrace{\frac{1}{\beta} \sum_{\omega\leqslant\nu/2}e^{-i\omega\tau}
K^{0\alpha\mathbf{k}}_{\lambda\lambda'}(s\mathbf{q};\omega;\nu)}_{smooth~function~of~\tau}
\nonumber\\&+e^{-i\nu\tau}\underbrace{\frac{1}{\beta}\sum_{\omega\geqslant\nu/2}e^{-i(\omega-\nu)\tau}
K^{0\alpha\mathbf{k}}_{\lambda\lambda'}(s\mathbf{q};\omega;\nu)}_{smooth~function~of~\tau}.
\end{align}

At this point, it is convenient to introduce two functions

\begin{align}\label{K0_7}
&K1^{0\alpha\mathbf{k}}_{\lambda\lambda'}(s\mathbf{q};\tau;\nu)\nonumber\\&=\Big\{\frac{1}{\beta} \sum_{\omega\geqslant\nu/2}e^{-i(\omega-\nu)\tau}
K^{^{*}0\alpha\mathbf{k}}_{\lambda\lambda'}(s\mathbf{q};-\omega+\nu;\nu)\Big\}^{*},
\end{align}
and
\begin{align}\label{K0_8}
K2^{0\alpha\mathbf{k}}_{\lambda\lambda'}(s\mathbf{q};\tau;\nu)=\frac{1}{\beta}\sum_{\omega\geqslant\nu/2}e^{-i(\omega-\nu)\tau}
K^{0\alpha\mathbf{k}}_{\lambda\lambda'}(s\mathbf{q};\omega;\nu).
\end{align}

Now the following Hartree-Fock contribution in $(\tau,\nu)$-representation which was subtracted earlier in $(\omega,\nu)$-representation is added (what is to be added to K1(K2) is clear from the structure of the formula (\ref{K0_9}))

\begin{align}\label{K0_9}
K^{0,HF,\alpha\mathbf{k}}_{\lambda\lambda'}&(s\mathbf{q};\tau;\nu)= \frac{
\langle\Psi^{\alpha\mathbf{k}}_{\lambda}|\Psi^{\alpha\mathbf{k}-\mathbf{q}}_{\lambda'}\Pi^{\mathbf{q}}_{s}\rangle}
{i\nu+\epsilon^{\alpha\mathbf{k}-\mathbf{q}}_{\lambda'}-\epsilon^{\alpha\mathbf{k}}_{\lambda}}\nonumber\\&\times
\Big\{G^{HF,\alpha,\mathbf{k}}_{\lambda}(\tau)-e^{-i\nu\tau}G^{HF,\alpha,\mathbf{k}-\mathbf{q}}_{\lambda'}(\tau)\Big\}.
\end{align}
In case $\nu=0$ and $\epsilon^{\alpha\mathbf{k}-\mathbf{q}}_{\lambda'}=\epsilon^{\alpha\mathbf{k}}_{\lambda}$ the expression is different

\begin{align}\label{K0_10}
K^{0,HF,\alpha\mathbf{k}}_{\lambda\lambda'}(s\mathbf{q};\tau;\nu)&=
\langle\Psi^{\alpha\mathbf{k}}_{\lambda}|\Psi^{\alpha\mathbf{k}-\mathbf{q}}_{\lambda'}\Pi^{\mathbf{q}}_{s}\rangle
G^{HF,\alpha,\mathbf{k}}_{\lambda}(\tau)\nonumber\\&\times\Big\{\tau+\beta G^{HF,\alpha,\mathbf{k}}_{\lambda}(\beta)\Big\}.
\end{align}

\subsection{ $K$-function in real space} \label{K_real}

Specific formula to be used to transform the $K$-function to the real space depends on where its two space arguments belong (MT-sphere or the
interstitial region). Correspondingly there are four different cases shown below:

Mt-Mt
\begin{align}\label{KR_1}
&K^{\alpha\mathbf{R}}_{\mathbf{t}L;\mathbf{t}'L'}(s\mathbf{q};\tau;\nu)=\nonumber\\&\frac{1}{N_{\mathbf{k}}}\sum_{\mathbf{k}}e^{i\mathbf{k}\mathbf{R}}
\sum_{\lambda\lambda'}Z^{\alpha\mathbf{k}}_{\mathbf{t}L;\lambda}K^{\alpha\mathbf{k}}_{\lambda\lambda'}(s\mathbf{q};\tau;\nu)
Z^{^{*}\alpha\mathbf{k}-\mathbf{q}}_{\mathbf{t}'L';\lambda'},
\end{align}

Int-Mt
\begin{align}\label{KR_2}
&K^{\alpha\mathbf{R}}_{\mathbf{r};\mathbf{t}'L'}(s\mathbf{q};\tau;\nu)=\nonumber\\&\frac{1}{N_{\mathbf{k}}}\sum_{\mathbf{k}}e^{i\mathbf{k}\mathbf{R}}
\sum_{\lambda\lambda'}A^{\alpha\mathbf{k}}_{\mathbf{r};\lambda}K^{\alpha\mathbf{k}}_{\lambda\lambda'}(s\mathbf{q};\tau;\nu)
Z^{^{*}\alpha\mathbf{k}-\mathbf{q}}_{\mathbf{t}'L';\lambda'},
\end{align}

Mt-Int
\begin{align}\label{KR_3}
&K^{\alpha\mathbf{R}}_{\mathbf{t}L;\mathbf{r}'}(s\mathbf{q};\tau;\nu)=\nonumber\\&\frac{1}{N_{\mathbf{k}}}\sum_{\mathbf{k}}e^{i\mathbf{k}\mathbf{R}}
\sum_{\lambda\lambda'}Z^{\alpha\mathbf{k}}_{\mathbf{t}L;\lambda}K^{\alpha\mathbf{k}}_{\lambda\lambda'}(s\mathbf{q};\tau;\nu)
A^{^{*}\alpha\mathbf{k}-\mathbf{q}}_{\mathbf{r}';\lambda'},
\end{align}

Int-Int
\begin{align}\label{KR_4}
K^{\alpha\mathbf{R}}_{\mathbf{r};\mathbf{r}'}&(s\mathbf{q};\tau;\nu)=\nonumber\\&\frac{1}{N_{\mathbf{k}}}\sum_{\mathbf{k}}e^{i\mathbf{k}\mathbf{R}}
\sum_{\lambda\lambda'}A^{\alpha\mathbf{k}}_{\mathbf{r};\lambda}K^{\alpha\mathbf{k}}_{\lambda\lambda'}(s\mathbf{q};\tau;\nu)
A^{^{*}\alpha\mathbf{k}-\mathbf{q}}_{\mathbf{r}';\lambda'},
\end{align}
with
\begin{align}\label{KR_5}
A^{\alpha\mathbf{k}}_{\mathbf{r};\lambda}=\frac{1}{\sqrt{\Omega_{0}}}\sum_{\mathbf{G}}e^{i(\mathbf{k}+\mathbf{G})\mathbf{r}}
A^{\alpha\mathbf{k}}_{\mathbf{G};\lambda}.
\end{align}

The coefficients $A^{\alpha\mathbf{k}}_{\mathbf{G};\lambda}$ represent the expansion of band states in plane waves in the interstitial region
$\Psi^{\alpha\mathbf{k}}_{\lambda}(\mathbf{r})=\frac{1}{\Omega_{0}}
\sum_{\mathbf{G}}A^{\alpha\mathbf{k}}_{\mathbf{G};\lambda}e^{i(\mathbf{k}+\mathbf{G})\mathbf{r}}$, and the coefficients
$Z^{\alpha\mathbf{k}}_{\mathbf{t}L;\lambda}$ represent the expansion of the band states in the orbital basis inside MT spheres
$\Psi^{\alpha\mathbf{k}}_{\lambda}(\mathbf{r})|_{\mathbf{t}}=
\sum_{L}Z^{\alpha\mathbf{k}}_{\mathbf{t}L;\lambda}\phi^{\alpha\mathbf{t}}_{L}(\mathbf{r})$.

\subsection{Evaluation of $W(21)K(123)$} \label{WK_calc}

The first term on the right hand side of the formula (\ref{d_Vert3}) can be rewritten with explicit $\tau$- and frequency-dependencies as the following

\begin{align}\label{dLam_WK_1}
\triangle \Gamma^{\alpha}(123;\tau;\nu)=W(12;\tau) K^{\alpha}(123;\tau;\nu).
\end{align}

For both $K1$- and $K2$-components one obtains the following formulae in the real space (distinguishing again MT and the interstitial region):

Mt-Mt
\begin{align}\label{dLam_WK_3}
&\triangle \Gamma^{\alpha\mathbf{R}}_{\mathbf{t}L;\mathbf{t}'L'}(s\mathbf{q};\tau;\nu)=\nonumber\\&\sum_{s'L''} \sum_{s''}\sum_{L'''}K^{\alpha\mathbf{R}}_{\mathbf{t}L'';\mathbf{t}'L'''}(s\mathbf{q};\tau;\nu)
\langle\phi^{\alpha\mathbf{t}'}_{L'}|\phi^{\alpha\mathbf{t}'}_{L'''}\Pi^{\mathbf{t}'}_{s''}\rangle^{*}\nonumber\\&\times
W^{\mathbf{R}}_{\mathbf{t}s';\mathbf{t}'s''}(\tau)\langle\phi^{\alpha\mathbf{t}}_{L}|\phi^{\alpha\mathbf{t}}_{L''}\Pi^{\mathbf{t}}_{s'}\rangle,
\end{align}

Int-Mt
\begin{align}\label{dLam_WK_4}
\triangle \Gamma^{\alpha\mathbf{R}}_{\mathbf{r};\mathbf{t}'L'}&(s\mathbf{q};\tau;\nu)=\sum_{L'''} \sum_{s''}\langle\phi^{\alpha\mathbf{t}'}_{L'}|\phi^{\alpha\mathbf{t}'}_{L'''}\Pi^{\mathbf{t}'}_{s''}\rangle^{*}
\nonumber\\&\times W^{\mathbf{R}}_{\mathbf{r};\mathbf{t}'s''}(\tau)K^{\alpha\mathbf{R}}_{\mathbf{r};\mathbf{t}'L'''}(s\mathbf{q};\tau;\nu),
\end{align}

Mt-Int
\begin{align}\label{dLam_WK_5}
\triangle \Gamma^{\alpha\mathbf{R}}_{\mathbf{t}L;\mathbf{r}'}(s\mathbf{q};\tau;\nu)&=\sum_{L''} \sum_{s'} \langle\phi^{\alpha\mathbf{t}}_{L}|\phi^{\alpha\mathbf{t}}_{L''}\Pi^{\mathbf{t}}_{s'}\rangle
W^{\mathbf{R}}_{\mathbf{t}s';\mathbf{r}'}(\tau) \nonumber\\&\times K^{\alpha\mathbf{R}}_{\mathbf{t}L'';\mathbf{r}'}(s\mathbf{q};\tau;\nu),
\end{align}

Int-Int
\begin{align}\label{dLam_WK_6}
\triangle \Gamma^{\alpha\mathbf{R}}_{\mathbf{r};\mathbf{r}'}(s\mathbf{q};\tau;\nu)= W^{\mathbf{R}}_{\mathbf{r};\mathbf{r}'}(\tau)
K^{\alpha\mathbf{R}}_{\mathbf{r};\mathbf{r}'}(s\mathbf{q};\tau;\nu).
\end{align}

In practical calculations one has to separate static and dynamic parts of the interaction $W=V+\widetilde{W}$. Correspondingly, static
and dynamic parts of the vertex correction are considered separately. Particularly, there is no $\tau$-dependence in the static part. Formulae (\ref{dLam_WK_3}-\ref{dLam_WK_6}) are the same for dynamic parts $\triangle\Gamma 1$ ($K1$ is used instead of $K$) and $\triangle\Gamma 2$ ($K2$ is used instead of $K$). For the static part $\triangle\Gamma^{stat}(\nu)$, one replaces $W(\tau)$ with $V$ and, correspondingly, $K1(\tau=0,\nu)+K2(\tau=0,\nu)$ is used instead of $K(\tau,\nu)$. 

Equation (\ref{dK_def}) can be
used most efficiently with quantities in band/frequency representation. Thus, $\triangle \Gamma1$, $\triangle \Gamma2$, and $\triangle\Gamma^{stat}$ are transformed into the band representation first

\begin{align}\label{dLam_WK_6a}
&\triangle \Gamma^{\alpha\mathbf{k}}_{\lambda\lambda'}(s\mathbf{q};\tau;\nu)=\nonumber\\&
\sum_{\mathbf{t}L}\sum_{\mathbf{t}'L'}Z^{^{*}\alpha\mathbf{k}\lambda}_{\mathbf{t}L}\sum_{\mathbf{R}}e^{-i\mathbf{k}\mathbf{R}}\triangle
\Gamma^{\alpha\mathbf{R}}_{\mathbf{t}L;\mathbf{t}'L'}(s\mathbf{q};\tau;\nu)Z^{\alpha\mathbf{k}-\mathbf{q}\lambda'}_{\mathbf{t}'L'}\nonumber\\&+
\sum_{\mathbf{r}}\sum_{\mathbf{t}'L'}X^{^{*}\alpha\mathbf{k}}_{\mathbf{r}\lambda}\sum_{\mathbf{R}}e^{-i\mathbf{k}\mathbf{R}}\triangle
\Gamma^{\alpha\mathbf{R}}_{\mathbf{r};\mathbf{t}'L'}(s\mathbf{q};\tau;\nu)Z^{\alpha\mathbf{k}-\mathbf{q}\lambda'}_{\mathbf{t}'L'}\nonumber\\&+
\sum_{\mathbf{t}L}\sum_{\mathbf{r}'}Z^{^{*}\alpha\mathbf{k}\lambda}_{\mathbf{t}L}\sum_{\mathbf{R}}e^{-i\mathbf{k}\mathbf{R}}\triangle
\Gamma^{\alpha\mathbf{R}}_{\mathbf{t}L;\mathbf{r}'}(s\mathbf{q};\tau;\nu)X^{\alpha\mathbf{k}-\mathbf{q}}_{\mathbf{r}'\lambda'}\nonumber\\&+
\sum_{\mathbf{r}}\sum_{\mathbf{r}'}X^{^{*}\alpha\mathbf{k}}_{\mathbf{r}\lambda}\sum_{\mathbf{R}}e^{-i\mathbf{k}\mathbf{R}}\triangle
\Gamma^{\alpha\mathbf{R}}_{\mathbf{r};\mathbf{r}'}(s\mathbf{q};\tau;\nu)X^{\alpha\mathbf{k}-\mathbf{q}}_{\mathbf{r}'\lambda'},
\end{align}
with
\begin{align}\label{x_def}
X^{\alpha\mathbf{k}}_{\mathbf{r}\lambda}=\frac{1}{N_{\mathbf{r}}}\sum_{\mathbf{G}}e^{i(\mathbf{k}+\mathbf{G})\mathbf{r}}\Big\{
\int_{\Omega_{Int}}d\mathbf{r}\Psi^{^{*}\alpha\mathbf{k}}_{\lambda}(\mathbf{r})e^{i(\mathbf{k}+\mathbf{G})\mathbf{r}}\Big\}^{*}.
\end{align}

Formula (\ref{dLam_WK_6a}) is used for $\triangle \Gamma1$, $\triangle \Gamma2$, and $\triangle\Gamma^{stat}$ with $\tau=0$ for the latter.

Then one transforms dynamic functions $\triangle \Gamma1^{\alpha\mathbf{k}}_{\lambda\lambda'}(s\mathbf{q};\tau;\nu)$ and $\triangle
\Gamma2^{\alpha\mathbf{k}}_{\lambda\lambda'}(s\mathbf{q};\tau;\nu)$ into $\triangle
\Gamma^{\alpha\mathbf{k}}_{\lambda\lambda'}(s\mathbf{q};\omega;\nu)$ and $\triangle
\Gamma^{\alpha\mathbf{k}}_{\lambda\lambda'}(s\mathbf{q};-\omega+\nu;\nu)$ using the formula (\ref{ww_4})

\begin{align}\label{dLam_WK_7}
&\triangle \Gamma^{\alpha\mathbf{k}}_{\lambda\lambda'}(s\mathbf{q};\omega;\nu)= \int_{0}^{\beta/2} d\tau\Big\{
\cos(\omega\tau)\nonumber\\&\times[\triangle \Gamma1^{\alpha\mathbf{k}}_{\lambda\lambda'}(s\mathbf{q};\tau;\nu)
-\triangle \Gamma1^{\alpha\mathbf{k}}_{\lambda\lambda'}(s\mathbf{q};\beta-\tau;\nu)]\nonumber\\
&+i\sin(\omega\tau)\nonumber\\&\times[\triangle \Gamma1^{\alpha\mathbf{k}}_{\lambda\lambda'}(s\mathbf{q};\tau;\nu)
+\triangle \Gamma1^{\alpha\mathbf{k}}_{\lambda\lambda'}(s\mathbf{q};\beta-\tau;\nu)]\nonumber\\
&+\cos((\omega-\nu)\tau)\nonumber\\&\times[\triangle \Gamma2^{\alpha\mathbf{k}}_{\lambda\lambda'}(s\mathbf{q};\tau;\nu)
-\triangle \Gamma2^{\alpha\mathbf{k}}_{\lambda\lambda'}(s\mathbf{q};\beta-\tau;\nu)]\nonumber\\
&+i\sin((\omega-\nu)\tau)\nonumber\\&\times[\triangle \Gamma2^{\alpha\mathbf{k}}_{\lambda\lambda'}(s\mathbf{q};\tau;\nu) +\triangle
\Gamma2^{0\alpha\mathbf{k}}_{\lambda\lambda'}(s\mathbf{q};\beta-\tau;\nu)]\Big\},
\end{align}
and
\begin{align}\label{dLam_WK_8}
&\triangle \Gamma^{\alpha\mathbf{k}}_{\lambda\lambda'}(s\mathbf{q};-\omega+\nu;\nu)= \int_{0}^{\beta/2} d\tau\Big\{
\cos(\omega\tau)\nonumber\\&\times[\triangle \Gamma2^{\alpha\mathbf{k}}_{\lambda\lambda'}(s\mathbf{q};\tau;\nu)
-\triangle \Gamma2^{\alpha\mathbf{k}}_{\lambda\lambda'}(s\mathbf{q};\beta-\tau;\nu)]\nonumber\\
&-i\sin(\omega\tau)\nonumber\\&\times[\triangle \Gamma2^{\alpha\mathbf{k}}_{\lambda\lambda'}(s\mathbf{q};\tau;\nu)
+\triangle \Gamma2^{\alpha\mathbf{k}}_{\lambda\lambda'}(s\mathbf{q};\beta-\tau;\nu)]\nonumber\\
&+\cos((\omega-\nu)\tau)\nonumber\\&\times[\triangle \Gamma1^{\alpha\mathbf{k}}_{\lambda\lambda'}(s\mathbf{q};\tau;\nu)
-\triangle \Gamma1^{\alpha\mathbf{k}}_{\lambda\lambda'}(s\mathbf{q};\beta-\tau;\nu)]\nonumber\\
&-i\sin((\omega-\nu)\tau)\nonumber\\&\times[\triangle \Gamma1^{\alpha\mathbf{k}}_{\lambda\lambda'}(s\mathbf{q};\tau;\nu) +\triangle
\Gamma1^{0\alpha\mathbf{k}}_{\lambda\lambda'}(s\mathbf{q};\beta-\tau;\nu)]\Big\}.
\end{align}

\subsection{ $\Delta K$ calculation} \label{dK_calc}

Equation (\ref{dK_def}) can be rewritten with explicit $\tau$-dependence as the following

\begin{align}\label{dK_1}
&\triangle K^{\alpha}(123;\tau;\tau')=-\int\int
d(45)d\tau''d\tau'''G^{\alpha}(14;\tau-\tau'')\nonumber\\&\times\triangle\Gamma^{\alpha}(453;\tau'';\tau''') G^{\alpha}(52;\tau'''-\tau'),
\end{align}
or, in frequency representation
\begin{align}\label{dK_2}
\triangle K^{\alpha}(123;\omega;\nu)&=-\int\int d(45)G^{\alpha}(14;\omega)\nonumber\\&\times\triangle\Gamma^{\alpha}(453;\omega;\nu)
G^{\alpha}(52;\omega-\nu).
\end{align}

It is also convenient to evaluate it in the band representation

\begin{align}\label{dK_3}
\triangle K^{0\alpha\mathbf{k}}_{\lambda\lambda'}&(s\mathbf{q};\omega;\nu)=
-\sum_{\lambda''\lambda'''}G^{\alpha\mathbf{k}}_{\lambda\lambda''}(\omega) \nonumber\\&\times\triangle
\Gamma^{\alpha\mathbf{k}}_{\lambda''\lambda'''}(s\mathbf{q};\omega;\nu) G^{\alpha,\mathbf{k}-\mathbf{q}}_{\lambda'''\lambda'}(\omega-\nu),
\end{align}
and
\begin{align}\label{dK_4}
\triangle K^{0\alpha\mathbf{k}}_{\lambda\lambda'}&(s\mathbf{q};-\omega+\nu;\nu)=
-\sum_{\lambda''\lambda'''}G^{^{*}\alpha\mathbf{k}}_{\lambda''\lambda}(\omega-\nu) \nonumber\\&\times\triangle
\Gamma^{\alpha\mathbf{k}}_{\lambda''\lambda'''}(s\mathbf{q};-\omega+\nu;\nu) G^{^{*}\alpha,\mathbf{k}-\mathbf{q}}_{\lambda'\lambda'''}(\omega).
\end{align}

The vertex in the Eqs. (\ref{dK_3}) and (\ref{dK_4}) represents the sum of dynamic $\triangle\Gamma(\omega,\nu)$ and static $\triangle\Gamma^{stat}(\nu)$ parts.

\subsection{$Q$ calculation} \label{Q_calc}

It is convenient to evaluate Eq.(\ref{Q_def}) in real space and $(\tau,\nu)$-representation. Considering again four cases according to the MT geometry, one obtains:

Mt-Mt
\begin{align}\label{Q_6}
&Q1^{\mathbf{R}}_{\mathbf{t}s';\mathbf{t}'s''}(s\mathbf{q};\tau;\nu)=\sum_{\alpha}\sum_{LL'}
\sum_{L''L'''}\nonumber\\&\Big\{\langle\phi^{\alpha\mathbf{t}}_{L''}|\phi^{\alpha\mathbf{t}}_{L}\Pi^{\mathbf{t}}_{s'}\rangle^{*}
G^{\alpha,-\mathbf{R}}_{\mathbf{t}'L';\mathbf{t}L}(-\tau)\nonumber\\&\times
\langle\phi^{\alpha\mathbf{t}'}_{L'''}|\phi^{\alpha\mathbf{t}'}_{L'}\Pi^{\mathbf{t}'}_{s''}\rangle
K1^{\alpha\mathbf{R}}_{\mathbf{t}L'';\mathbf{t}'L'''}(s\mathbf{q};\tau;\nu)\nonumber\\&+
e^{i\mathbf{q}\mathbf{R}}\langle\phi^{\alpha\mathbf{t}}_{L''}|\phi^{\alpha\mathbf{t}}_{L}\Pi^{\mathbf{t}}_{s'}\rangle^{*}
G^{\alpha\mathbf{R}}_{\mathbf{t}L;\mathbf{t}'L'}(\tau)\nonumber\\&\times
\langle\phi^{\alpha\mathbf{t}'}_{L'''}|\phi^{\alpha\mathbf{t}'}_{L'}\Pi^{\mathbf{t}'}_{s''}\rangle
K2^{\alpha,-\mathbf{R}}_{\mathbf{t}'L''';\mathbf{t}L''}(s\mathbf{q};-\tau;\nu) \Big\},
\end{align}

Mt-Int
\begin{align}\label{Q_7}
&Q1^{\mathbf{R}}_{\mathbf{t}s';\mathbf{r}'}(s\mathbf{q};\tau;\nu)=\nonumber\\&\sum_{\alpha}\sum_{LL''}
\Big\{\langle\phi^{\alpha\mathbf{t}}_{L''}|\phi^{\alpha\mathbf{t}}_{L}\Pi^{\mathbf{t}}_{s'}\rangle^{*}
G^{\alpha,-\mathbf{R}}_{\mathbf{r}';\mathbf{t}L}(-\tau)K1^{\alpha\mathbf{R}}_{\mathbf{t}L'';\mathbf{r}'}(s\mathbf{q};\tau;\nu)\nonumber\\&+
e^{i\mathbf{q}\mathbf{R}}\langle\phi^{\alpha\mathbf{t}}_{L}|\phi^{\alpha\mathbf{t}}_{L''}\Pi^{\mathbf{t}}_{s'}\rangle^{*}
G^{\alpha\mathbf{R}}_{\mathbf{t}L;\mathbf{r}'}(\tau)K2^{\alpha,-\mathbf{R}}_{\mathbf{r}';\mathbf{t}L''}(s\mathbf{q};-\tau;\nu) \Big\},
\end{align}

Int-Mt
\begin{align}\label{Q_8}
&Q1^{\mathbf{R}}_{\mathbf{r};\mathbf{t}'s''}(s\mathbf{q};\tau;\nu)=\sum_{\alpha}\sum_{L'L'''}\nonumber\\
&\Big\{\langle\phi^{\alpha\mathbf{t}'}_{L'''}|\phi^{\alpha\mathbf{t}'}_{L'}\Pi^{\mathbf{t}'}_{s''}\rangle
G^{\alpha,-\mathbf{R}}_{\mathbf{t}'L';\mathbf{r}}(-\tau)K1^{\alpha\mathbf{R}}_{\mathbf{r};\mathbf{t}'L'''}(s\mathbf{q};\tau;\nu)\nonumber\\&+
e^{i\mathbf{q}\mathbf{R}}\langle\phi^{\alpha\mathbf{t}'}_{L'}|\phi^{\alpha\mathbf{t}'}_{L'''}\Pi^{\mathbf{t}'}_{s''}\rangle
G^{\alpha\mathbf{R}}_{\mathbf{r};\mathbf{t}'L'}(\tau)K2^{\alpha,-\mathbf{R}}_{\mathbf{t}'L''';\mathbf{r}}(s\mathbf{q};-\tau;\nu) \Big\},
\end{align}

Int-Int
\begin{align}\label{Q_9}
Q1^{\mathbf{R}}_{\mathbf{r};\mathbf{r}'}&(s\mathbf{q};\tau;\nu)=\sum_{\alpha}\Big\{
G^{\alpha,-\mathbf{R}}_{\mathbf{r}';\mathbf{r}}(-\tau)K1^{\alpha\mathbf{R}}_{\mathbf{r};\mathbf{r}'}(s\mathbf{q};\tau;\nu)\nonumber\\&+
e^{i\mathbf{q}\mathbf{R}}
G^{\alpha\mathbf{R}}_{\mathbf{r};\mathbf{r}'}(\tau)K2^{\alpha,-\mathbf{R}}_{\mathbf{r}';\mathbf{r}}(s\mathbf{q};-\tau;\nu) \Big\}.
\end{align}

To evaluate $Q2$, the same expression is used with replacement $K1\leftrightarrow K2$.
Then follows the transformation to the $\mathbf{q}$-space:

Mt-Mt
\begin{align}\label{Q_10}
Q^{\mathbf{q}'}_{\mathbf{t}s';\mathbf{t}'s''}(s\mathbf{q};\tau;\nu)=\sum_{\mathbf{R}}e^{-i\mathbf{q}'\mathbf{R}}
Q^{\mathbf{R}}_{\mathbf{t}s';\mathbf{t}'s''}(s\mathbf{q};\tau;\nu),
\end{align}
Int-Mt
\begin{align}\label{Q_11}
Q^{\mathbf{q}'}_{s';\mathbf{t}'s''}(s\mathbf{q};\tau;\nu)=\sum_{\mathbf{r}}Y^{^{*}\mathbf{q}'}_{\mathbf{r};s'}
\sum_{\mathbf{R}}e^{-i\mathbf{q}'\mathbf{R}} Q^{\mathbf{R}}_{\mathbf{r};\mathbf{t}'s''}(s\mathbf{q};\tau;\nu),
\end{align}
Mt-Int
\begin{align}\label{Q_12}
Q^{\mathbf{q}'}_{\mathbf{t}s';s''}(s\mathbf{q};\tau;\nu)=\sum_{\mathbf{r}'}Y^{\mathbf{q}'}_{\mathbf{r}';s''}
\sum_{\mathbf{R}}e^{-i\mathbf{q}'\mathbf{R}} Q^{\mathbf{R}}_{\mathbf{t}s';\mathbf{r}'}(s\mathbf{q};\tau;\nu),
\end{align}
Int-Int
\begin{align}\label{Q_13}
Q^{\mathbf{q}'}_{s';s''}(s\mathbf{q};\tau;\nu)=\sum_{\mathbf{r}\mathbf{r}'} Y^{^{*}\mathbf{q}'}_{\mathbf{r};s'}Y^{\mathbf{q}'}_{\mathbf{r}';s''}
\sum_{\mathbf{R}}e^{-i\mathbf{q}'\mathbf{R}} Q^{\mathbf{R}}_{\mathbf{r};\mathbf{r}'}(s\mathbf{q};\tau;\nu),
\end{align}
with
\begin{align}\label{y_def}
Y^{\mathbf{q}}_{\mathbf{r};s}=\frac{1}{N_{\mathbf{r}}}\sum_{\mathbf{G}}e^{i(\mathbf{q}+\mathbf{G})\mathbf{r}}
\int_{\Omega_{Int}}d\mathbf{r}e^{i(\mathbf{G}-\mathbf{G}_{s})\mathbf{r}}.
\end{align}

Finally, $Q$-function is transformed to $(\nu',\nu)$-representation in order to be used in $T$-evaluation (next subsection):

\begin{align}\label{Q_14}
&Q^{\mathbf{q}'}_{ns's''}(s\mathbf{q};\nu';\nu)=\int_{0}^{\beta/2}d\tau \Big\{
\cos(\nu'\tau)\nonumber\\&\times\big[Q1^{\mathbf{q}'}_{s's''}(s\mathbf{q};\tau;\nu)+
Q1^{\mathbf{q}'}_{s's''}(s\mathbf{q};-\tau;\nu)\big]\nonumber\\&
+i\sin(\nu'\tau)\nonumber\\&\times\big[Q1^{\mathbf{q}'}_{s's''}(s\mathbf{q};\tau;\nu)-
Q1^{\mathbf{q}'}_{s's''}(s\mathbf{q};-\tau;\nu)\big]\nonumber\\&
+\cos(\nu'-\nu)\tau\nonumber\\&\times\big[Q2^{\mathbf{q}'}_{s's''}(s\mathbf{q};\tau;\nu)+
Q2^{\mathbf{q}'}_{s's''}(s\mathbf{q};-\tau;\nu)\big]\nonumber\\&
+i\sin(\nu'-\nu)\tau\nonumber\\&\times\big[Q2^{\mathbf{q}'}_{s's''}(s\mathbf{q};\tau;\nu)-
Q2^{\mathbf{q}'}_{s's''}(s\mathbf{q};-\tau;\nu)\big]\Big\},
\end{align}
and
\begin{align}\label{Q_15}
&Q^{\mathbf{q}'}_{ns's''}(s\mathbf{q};-\nu'+\nu;\nu)=\int_{0}^{\beta/2}d\tau \Big\{
\cos(\nu'\tau)\nonumber\\&\times\big[Q2^{\mathbf{q}'}_{s's''}(s\mathbf{q};\tau;\nu)+
Q2^{\mathbf{q}'}_{s's''}(s\mathbf{q};-\tau;\nu)\big]\nonumber\\&
-i\sin(\nu'\tau)\nonumber\\&\times\big[Q2^{\mathbf{q}'}_{s's''}(s\mathbf{q};\tau;\nu)-
Q2^{\mathbf{q}'}_{s's''}(s\mathbf{q};-\tau;\nu)\big]\nonumber\\&
+\cos(\nu'-\nu)\tau\nonumber\\&\times\big[Q1^{\mathbf{q}'}_{s's''}(s\mathbf{q};\tau;\nu)+
Q1^{\mathbf{q}'}_{s's''}(s\mathbf{q};-\tau;\nu)\big]\nonumber\\&
-i\sin(\nu'-\nu)\tau\nonumber\\&\times\big[Q1^{\mathbf{q}'}_{s's''}(s\mathbf{q};\tau;\nu)-
Q1^{\mathbf{q}'}_{s's''}(s\mathbf{q};-\tau;\nu)\big]\Big\}.
\end{align}

\subsection{ $T$ calculation} \label{T_calc}

From the equation (\ref{T_def}) one obtains the T-function in the RPB representation
\begin{align}\label{T_12}
T^{\mathbf{q}'}_{s's''}&(s\mathbf{q};\nu';\nu)=\sum_{s_{1}s_{2}} W^{\mathbf{q}'}_{s's_{1}}(\nu')\nonumber\\&\times
Q^{\mathbf{q}'}_{s_{1}s_{2}}(s\mathbf{q};\nu';\nu) W^{\mathbf{q}'-\mathbf{q}}_{s_{2}s''}(\nu'-\nu),
\end{align}
and
\begin{align}\label{T_13}
T^{\mathbf{q}'}_{s's''}&(s\mathbf{q};-\nu'+\nu;\nu)=\sum_{s_{1}s_{2}} W^{\mathbf{q}'}_{s's_{1}}(\nu'-\nu)\nonumber\\&\times
Q^{\mathbf{q}'}_{s_{1}s_{2}}(s\mathbf{q};-\nu'+\nu;\nu) W^{\mathbf{q}'-\mathbf{q}}_{s_{2}s''}(\nu').
\end{align}

In order to perform the transform $\nu'\rightarrow\tau$ the static contribution is subtracted first

\begin{align}\label{T_12a}
T^{\mathbf{q}'}_{s's''}(s\mathbf{q};\nu';\nu)&=T^{\mathbf{q}'}_{s's''}(s\mathbf{q};\nu';\nu)\nonumber\\&-\sum_{s_{1}s_{2}}
V^{\mathbf{q}'}_{s's_{1}} Q^{\mathbf{q}'}_{s_{1}s_{2}}(s\mathbf{q};\nu';\nu) V^{\mathbf{q}'-\mathbf{q}}_{s_{2}s''},
\end{align}
and
\begin{align}\label{T_13a}
T^{\mathbf{q}'}_{s's''}&(s\mathbf{q};-\nu'+\nu;\nu)=T^{\mathbf{q}'}_{s's''}(s\mathbf{q};-\nu'+\nu;\nu)\nonumber\\&-\sum_{s_{1}s_{2}}
V^{\mathbf{q}'}_{s's_{1}} Q^{\mathbf{q}'}_{s_{1}s_{2}}(s\mathbf{q};-\nu'+\nu;\nu) V^{\mathbf{q}'-\mathbf{q}}_{s_{2}s''}.
\end{align}

After that the transformation is accomplished straightforwardly:

\begin{align}\label{T_14}
T^{\mathbf{q}'}_{s's''}&(s\mathbf{q};\tau;\nu)=\frac{1}{\beta}\sum_{\nu'\leqslant\nu/2} e^{-i\nu'\tau}
T^{\mathbf{q}'}_{s's''}(s\mathbf{q};\nu';\nu)\nonumber\\&+\frac{1}{\beta}\sum_{\nu'\geqslant\nu/2} e^{-i\nu'\tau}
T^{\mathbf{q}'}_{s's''}(s\mathbf{q};\nu';\nu)\nonumber\\&= T1^{\mathbf{q}'}_{s's''}(s\mathbf{q};\tau;\nu)
+e^{-i\nu\tau}T2^{\mathbf{q}'}_{s's''}(s\mathbf{q};\tau;\nu),
\end{align}
where the following notations have been defined

\begin{align}\label{T_15}
T1^{\mathbf{q}'}_{s's''}&(s\mathbf{q};\tau;\nu)=\frac{1}{\beta}\sum_{\nu'\leqslant\nu/2} e^{-i\nu'\tau}
T^{\mathbf{q}'}_{s's''}(s\mathbf{q};\nu';\nu)\nonumber\\&= \frac{1}{\beta}\sum_{\nu'\geqslant\nu/2} e^{i(\nu'-\nu)\tau}
T^{\mathbf{q}'}_{s's''}(s\mathbf{q};-\nu'+\nu;\nu)\nonumber\\&= \Big\{\frac{1}{\beta}\sum_{\nu'\geqslant\nu/2} e^{-i(\nu'-\nu)\tau}
T^{^{*}\mathbf{q}'}_{s's''}(s\mathbf{q};-\nu'+\nu;\nu)\Big\}^{*},
\end{align}
and
\begin{align}\label{T_16}
T2^{\mathbf{q}'}_{s's''}(s\mathbf{q};\tau;\nu)=\frac{1}{\beta}\sum_{\nu'\geqslant\nu/2} e^{-i(\nu'-\nu)\tau}
T^{\mathbf{q}'}_{s's''}(s\mathbf{q};\nu';\nu).
\end{align}

Now one adds the static contribution which was subtracted before
\begin{align}\label{T_16a}
T2^{\mathbf{q}'}_{s's''}&(s\mathbf{q};\tau;\nu)=T2^{\mathbf{q}'}_{s's''}(s\mathbf{q};\tau;\nu)\nonumber\\&+\sum_{s_{1}s_{2}}
V^{\mathbf{q}'}_{s's_{1}} Q2^{\mathbf{q}'}_{s_{1}s_{2}}(s\mathbf{q};\tau;\nu) V^{\mathbf{q}'-\mathbf{q}}_{s_{2}s''},
\end{align}
and
\begin{align}\label{T_16b}
T1^{\mathbf{q}'}_{s's''}&(s\mathbf{q};\tau;\nu)=T1^{\mathbf{q}'}_{s's''}(s\mathbf{q};\tau;\nu)\nonumber\\&+\sum_{s_{1}s_{2}}
V^{\mathbf{q}'}_{s's_{1}} Q1^{\mathbf{q}'}_{s_{1}s_{2}}(s\mathbf{q};\tau;\nu) V^{\mathbf{q}'-\mathbf{q}}_{s_{2}s''}.
\end{align}

Finally, the $T$-function is transformed in the real space-representation to be used in (\ref{d_Vert3}):

Mt-Mt
\begin{align}\label{T_17}
T^{\mathbf{R}}_{\mathbf{t}s';\mathbf{t}'s''}(s\mathbf{q};\tau;\nu)=\frac{1}{N_{\mathbf{k}}}\sum_{\mathbf{q}'}e^{i\mathbf{q}'\mathbf{R}}
T^{\mathbf{q}'}_{\mathbf{t}s';\mathbf{t}'s''}(s\mathbf{q};\tau;\nu),
\end{align}
Int-Mt
\begin{align}\label{T_18}
T^{\mathbf{R}}_{\mathbf{r};\mathbf{t}'s''}(s\mathbf{q};\tau;\nu)=\frac{1}{N_{\mathbf{k}}}\sum_{\mathbf{q}'}e^{i\mathbf{q}'\mathbf{R}}
\sum_{s'}B^{\mathbf{q}'}_{\mathbf{r};s'} T^{\mathbf{q}'}_{s';\mathbf{t}'s''}(s\mathbf{q};\tau;\nu),
\end{align}
Mt-Int
\begin{align}\label{T_19}
T^{\mathbf{R}}_{\mathbf{t}s';\mathbf{r}'}(s\mathbf{q};\tau;\nu)=\frac{1}{N_{\mathbf{k}}}\sum_{\mathbf{q}'}e^{i\mathbf{q}'\mathbf{R}}
\sum_{s''}B^{^{*}\mathbf{q}'}_{\mathbf{r}';s''} T^{\mathbf{q}'}_{\mathbf{t}s';s''}(s\mathbf{q};\tau;\nu),
\end{align}
Int-Int
\begin{align}\label{T_20}
T^{\mathbf{R}}_{\mathbf{r};\mathbf{r}'}(s\mathbf{q};\tau;\nu)=\frac{1}{N_{\mathbf{k}}}\sum_{\mathbf{q}'}e^{i\mathbf{q}'\mathbf{R}} \sum_{s's''}
B^{\mathbf{q}'}_{\mathbf{r};s'}B^{^{*}\mathbf{q}'}_{\mathbf{r}';s''} T^{\mathbf{q}'}_{s's''}(s\mathbf{q};\tau;\nu),
\end{align}
with
\begin{align}\label{T_21}
B^{\mathbf{q}}_{\mathbf{r};s'}=e^{i(\mathbf{q}+\mathbf{G}_{s'})\mathbf{r}}.
\end{align}

\subsection{ $G(12)T(213)$ calculation} \label{GT_calc}

Second term on the rhs of (\ref{d_Vert3}) is evaluated in the real space and $(\tau,\nu)$-representation. Again, there are four different cases according to the MT-geometry:

Mt-Mt
\begin{align}\label{GT_4}
&\triangle \Gamma1^{\alpha\mathbf{R}}_{\mathbf{t}L;\mathbf{t}'L'}(s\mathbf{q};\tau;\nu)=e^{i\mathbf{q}\mathbf{R}}\sum_{L''L'''}
G^{\alpha\mathbf{R}}_{\mathbf{t}L'';\mathbf{t}'L'''}(\tau)\times\nonumber\\&
\sum_{s's''}\langle\phi^{\alpha\mathbf{t}}_{L''}|\phi^{\alpha\mathbf{t}}_{L}\Pi^{\mathbf{t}}_{s'}\rangle^{*}
T2^{-\mathbf{R}}_{\mathbf{t}'s'';\mathbf{t}s'}(s\mathbf{q};-\tau;\nu)
\langle\phi^{\alpha\mathbf{t}'}_{L'''}|\phi^{\alpha\mathbf{t}'}_{L'}\Pi^{\mathbf{t}'}_{s''}\rangle,
\end{align}

Mt-Int
\begin{align}\label{GT_5}
&\triangle \Gamma1^{\alpha\mathbf{R}}_{\mathbf{r};\mathbf{t}'L'}(s\mathbf{q};\tau;\nu)=e^{i\mathbf{q}\mathbf{R}}\sum_{L'''}
G^{\alpha\mathbf{R}}_{\mathbf{r};\mathbf{t}'L'''}(\tau) \nonumber\\& \sum_{s''}
T2^{-\mathbf{R}}_{\mathbf{t}'s'';\mathbf{r}}(s\mathbf{q};-\tau;\nu)
\langle\phi^{\alpha\mathbf{t}'}_{L'''}|\phi^{\alpha{\mathbf{t}'}_L'}\Pi^{\mathbf{t}'}_{s''}\rangle,
\end{align}

Int-Mt
\begin{align}\label{GT_6}
&\triangle \Gamma1^{\alpha\mathbf{R}}_{\mathbf{t}L;\mathbf{r}'}(s\mathbf{q};\tau;\nu)=e^{i\mathbf{q}\mathbf{R}}\sum_{L''}
G^{\alpha\mathbf{R}}_{\mathbf{t}L'';\mathbf{r}'}(\tau)\nonumber\\&
\sum_{s'}\langle\phi^{\alpha\mathbf{t}}_{L''}|\phi^{\alpha\mathbf{t}}_{L}\Pi^{\mathbf{t}}_{s'}\rangle^{*}
T2^{-\mathbf{R}}_{\mathbf{r}';\mathbf{t}s'}(s\mathbf{q};-\tau;\nu),
\end{align}

Int-Int
\begin{align}\label{GT_7}
\triangle \Gamma1^{\alpha\mathbf{R}}_{\mathbf{r};\mathbf{r}'}(s\mathbf{q};\tau;\nu)=e^{i\mathbf{q}\mathbf{R}}
G^{\alpha\mathbf{R}}_{\mathbf{r};\mathbf{r}'}(\tau) T2^{-\mathbf{R}}_{\mathbf{r}';\mathbf{r}}(s\mathbf{q};-\tau;\nu).
\end{align}

$\triangle\Gamma2$ is evaluated similarly with the replacement $T2\rightarrow T1$ in the formulae above.

\section{Correction to the Polarizability} \label{P_corr}

According to the Eqs. (\ref{def_pol1}) and (\ref{dK_def}) the correction to the polarizability can be written as the following

\begin{align}\label{p_corr_0}
\triangle P(12)=\sum_{\alpha}G^{\alpha}(13)\triangle\Gamma^{\alpha}(342)G^{\alpha}(41)=-\sum_{\alpha}\triangle K^{\alpha}(112).
\end{align}

It is represented in the RPB

\begin{align}\label{p_corr_1}
\triangle P(12;\nu)=\frac{1}{N_{\mathbf{k}}}\sum_{\mathbf{q}}\sum_{ss'}\widetilde{\Pi}^{\mathbf{q}}_{s}(1)\triangle
P^{\mathbf{q}}_{ss'}(\nu)\widetilde{\Pi}^{^{*}\mathbf{q}}_{s'}(2),
\end{align}
where the coefficients are found from the band representation of $\triangle K$
\begin{align}\label{p_corr_2}
\triangle P^{\mathbf{q}}_{ss'}(\nu)&=-\frac{1}{N_{\mathbf{k}}}\sum_{\alpha\mathbf{k}}\sum_{\lambda\lambda'} \langle
\Psi^{\alpha\mathbf{k}}_{\lambda}|\Psi^{\alpha\mathbf{k}-\mathbf{q}}_{\lambda'}\Pi^{\mathbf{q}}_{s}\rangle^{*}\nonumber\\& \times \triangle
K^{\alpha\mathbf{k}}_{\lambda\lambda'}(s'\mathbf{q};\tau=0;\nu).
\end{align}

After that the correction expressed in the full product basis ($M_{i(j)}$) can be found
\begin{align}\label{p_corr_3}
\triangle P^{\mathbf{q}}_{ij}(\nu)=\sum_{ss'}\langle M_{i}|\Pi_{s}\rangle\triangle P^{\mathbf{q}}_{ss'}(\nu)\langle\Pi_{s'}|M_{j}\rangle.
\end{align}

\section{Correction to the Self Energy} \label{Sig_corr}

In order to find the correction to the self energy, one can use general expression

\begin{align}\label{sig_1}
\triangle\Sigma^{\alpha}(12)=-G^{\alpha}(13)\triangle\Gamma^{\alpha}(324)W(41),
\end{align}
and, according to the separation of the vertex into dynamic $\triangle\Gamma^{dyn}=\triangle\Gamma(\omega,\nu)$ and static $\triangle\Gamma^{stat}(\nu)$ parts, and the separation of the screened interaction into Coulomb $V$ and dynamic $\widetilde{W}$ parts, one can divide the correction to the self-energy into dynamic, semi-dynamic, and static. They are considered below in this section.

In all cases the non-symmetrized self energy $\triangle\widetilde{\Sigma}$ is evaluated first. It is obtained when the summation runs only over irreducible $\mathbf{q}$-points with weights $w_{\mathbf{q}}$. In the end, the correction to the self energy is obtained according to the symmetrization procedure

\begin{align}\label{symmetr}
\triangle\Sigma^{\alpha\mathbf{k}}(\mathbf{r},\mathbf{r}';\tau)=\frac{1}{N_{A}}\sum_{A}\triangle\widetilde{\Sigma}^{\alpha A^{-1}\mathbf{k}}(A^{-1}\mathbf{r},A^{-1}\mathbf{r}';\tau),
\end{align}
where $A$ represents the symmetry operation, and $N_{A}$ is the number of symmetry operations.

\subsection{Correction to the Dynamic Self Energy} \label{Sig_corr_dyn}

The formulae of this subsection are applied when Eq.(\ref{sig_1}) is used with dynamic vertex and dynamic part of the interaction $\widetilde{W}$. In this case the expression (\ref{sig_1}) reads as the following

\begin{align}\label{sig_2}
&\triangle\Sigma^{dyn,\alpha}(12;\tau)=-\frac{1}{\beta}\sum_{\nu}e^{i\nu\tau}\frac{1}{\beta}\sum_{\omega}e^{-i\omega\tau}\nonumber\\& \times\int
d(34)G^{\alpha}(13;\omega)\triangle\Gamma^{\alpha}(324;\omega;\nu)\widetilde{W}(41;\nu),
\end{align}
where digits are used as space coordinates.

Introducing $\triangle\widetilde{\Sigma}^{\alpha\mathbf{k}}_{12}(\tau;\nu)$ through the relation

\begin{align}\label{sig_3}
\triangle\widetilde{\Sigma}^{dyn,\alpha\mathbf{k}}_{12}(\tau)=-\frac{1}{\beta}\sum_{\nu}e^{i\nu\tau}\triangle\widetilde{\Sigma}^{\alpha\mathbf{k}}_{12}(\tau;\nu),
\end{align}
one obtains

\begin{align}\label{sig_4}
\triangle\widetilde{\Sigma}^{\alpha\mathbf{k}}_{12}(\tau;\nu)=\frac{1}{\beta}\sum_{\omega}&e^{-i\omega\tau}\sum_{\mathbf{q}}w_{\mathbf{q}}
\sum_{s\lambda}G^{\alpha\mathbf{k}+\mathbf{q}}_{1;\lambda}(\omega)\nonumber\\&\times\triangle\Gamma^{\alpha\mathbf{k}+\mathbf{q}}_{\lambda;2}(s\mathbf{q};\omega;\nu)
\widetilde{W}^{\mathbf{q}}_{s;1}(\nu).
\end{align}

For the different locations of arguments $1$ and $2$ one gets:

Mt-Mt
\begin{align}\label{sig_5}
&\triangle\widetilde{\Sigma}^{\alpha\mathbf{k}}_{\mathbf{t}L;\mathbf{t}'L'}(\tau;\nu)=\sum_{\mathbf{q}}w_{\mathbf{q}}
\sum_{s}\sum_{L''}\nonumber\\&\Big\{\frac{1}{\beta}\sum_{\omega}e^{-i\omega\tau} \sum_{\lambda}
G^{\alpha\mathbf{k}+\mathbf{q}}_{\mathbf{t}L'';\lambda}(\omega)
\triangle\Gamma^{\alpha\mathbf{k}+\mathbf{q}}_{\lambda;\mathbf{t}'L'}(s\mathbf{q};\omega;\nu)\Big\}\nonumber\\&\times
\sum_{s'}\widetilde{W}^{\mathbf{q}}_{s;\mathbf{t}s'}(\nu)
\langle\phi^{\alpha\mathbf{t}}_{L''}|\phi^{\alpha\mathbf{t}}_{L}\Pi^{\mathbf{t}}_{s'}\rangle^{*}\nonumber\\&=\sum_{\mathbf{q}}w_{\mathbf{q}}
\sum_{s}\sum_{L''}\Big\{A1^{\alpha\mathbf{k}}_{\mathbf{t}L'';\mathbf{t}'L'}(s\mathbf{q};\tau;\nu)\nonumber\\&
+e^{-i\nu\tau}A2^{\alpha\mathbf{k}}_{\mathbf{t}L'';\mathbf{t}'L'}(s\mathbf{q};\tau;\nu)\Big\}
\widetilde{W}^{\alpha\mathbf{q}}_{s;\mathbf{t}L''L}(\nu)\nonumber\\&=\sum_{\mathbf{q}}w_{\mathbf{q}}
\sum_{s}\Big\{B1^{\alpha\mathbf{k}}_{\mathbf{t}L;\mathbf{t}'L'}(s\mathbf{q};\tau;\nu)\nonumber\\&
+e^{-i\nu\tau}B2^{\alpha\mathbf{k}}_{\mathbf{t}L;\mathbf{t}'L'}(s\mathbf{q};\tau;\nu)\Big\}\nonumber\\&=
C1^{\alpha\mathbf{k}}_{\mathbf{t}L;\mathbf{t}'L'}(\tau;\nu) +e^{-i\nu\tau}C2^{\alpha\mathbf{k}}_{\mathbf{t}L;\mathbf{t}'L'}(\tau;\nu),
\end{align}
with obvious notations,

Int-Mt
\begin{align}\label{sig_6}
&\triangle\widetilde{\Sigma}^{\alpha\mathbf{k}}_{\mathbf{r};\mathbf{t}'L'}(\tau;\nu)=\sum_{\mathbf{q}}w_{\mathbf{q}}
\sum_{s}\nonumber\\&\Big\{\frac{1}{\beta}\sum_{\omega}e^{-i\omega\tau} \sum_{\lambda} G^{\alpha\mathbf{k}+\mathbf{q}}_{\mathbf{r};\lambda}(\omega)
\triangle\Gamma^{\alpha\mathbf{k}+\mathbf{q}}_{\lambda;\mathbf{t}'L'}(s\mathbf{q};\omega;\nu)\Big\}\widetilde{W}^{\mathbf{q}}_{s;\mathbf{r}}(\nu)
\nonumber\\&=\sum_{\mathbf{q}}w_{\mathbf{q}} \sum_{s}\nonumber\\&\Big\{A1^{\alpha\mathbf{k}}_{\mathbf{r};\mathbf{t}'L'}(s\mathbf{q};\tau;\nu)
+e^{-i\nu\tau}A2^{\alpha\mathbf{k}}_{\mathbf{r};\mathbf{t}'L'}(s\mathbf{q};\tau;\nu)\Big\}
\widetilde{W}^{\alpha\mathbf{q}}_{s;\mathbf{r}}(\nu)\nonumber\\&=\sum_{\mathbf{q}}w_{\mathbf{q}}
\sum_{s}\Big\{B1^{\alpha\mathbf{k}}_{\mathbf{r};\mathbf{t}'L'}(s\mathbf{q};\tau;\nu)
+e^{-i\nu\tau}B2^{\alpha\mathbf{k}}_{\mathbf{r};\mathbf{t}'L'}(s\mathbf{q};\tau;\nu)\Big\}\nonumber\\&=
C1^{\alpha\mathbf{k}}_{\mathbf{r};\mathbf{t}'L'}(\tau;\nu) +e^{-i\nu\tau}C2^{\alpha\mathbf{k}}_{\mathbf{r};\mathbf{t}'L'}(\tau;\nu),
\end{align}

Mt-Int
\begin{align}\label{sig_7}
&\triangle\widetilde{\Sigma}^{\alpha\mathbf{k}}_{\mathbf{t}L;\mathbf{r}'}(\tau;\nu)=\sum_{\mathbf{q}}w_{\mathbf{q}}
\sum_{s}\sum_{L''}\nonumber\\&\Big\{\frac{1}{\beta}\sum_{\omega}e^{-i\omega\tau} \sum_{\lambda}
G^{\alpha\mathbf{k}+\mathbf{q}}_{\mathbf{t}L'';\lambda}(\omega)
\triangle\Gamma^{\alpha\mathbf{k}+\mathbf{q}}_{\lambda;\mathbf{r}'}(s\mathbf{q};\omega;\nu)\Big\}\nonumber\\&\times
\sum_{s'}\widetilde{W}^{\mathbf{q}}_{s;\mathbf{t}s'}(\nu)
\langle\phi^{\alpha\mathbf{t}}_{L''}|\phi^{\alpha\mathbf{t}}_{L}\Pi^{\mathbf{t}}_{s'}\rangle^{*}\nonumber\\&=\sum_{\mathbf{q}}w_{\mathbf{q}}
\sum_{s}\sum_{L''}\nonumber\\&\Big\{A1^{\alpha\mathbf{k}}_{\mathbf{t}L'';\mathbf{r}'}(s\mathbf{q};\tau;\nu)
+e^{-i\nu\tau}A2^{\alpha\mathbf{k}}_{\mathbf{t}L'';\mathbf{r}'}(s\mathbf{q};\tau;\nu)\Big\}
\widetilde{W}^{\alpha\mathbf{q}}_{s;\mathbf{t}L''L}(\nu)\nonumber\\&=\sum_{\mathbf{q}}w_{\mathbf{q}}
\sum_{s}\Big\{B1^{\alpha\mathbf{k}}_{\mathbf{t}L;\mathbf{r}'}(s\mathbf{q};\tau;\nu)
+e^{-i\nu\tau}B2^{\alpha\mathbf{k}}_{\mathbf{t}L;\mathbf{r}'}(s\mathbf{q};\tau;\nu)\Big\}\nonumber\\&=
C1^{\alpha\mathbf{k}}_{\mathbf{t}L;\mathbf{r}'}(\tau;\nu) +e^{-i\nu\tau}C2^{\alpha\mathbf{k}}_{\mathbf{t}L;\mathbf{r}'}(\tau;\nu),
\end{align}

Int-Int
\begin{align}\label{sig_8}
&\triangle\widetilde{\Sigma}^{\alpha\mathbf{k}}_{\mathbf{r};\mathbf{r}'}(\tau;\nu)=\sum_{\mathbf{q}}w_{\mathbf{q}}
\sum_{s}\nonumber\\&\Big\{\frac{1}{\beta}\sum_{\omega}e^{-i\omega\tau} \sum_{\lambda} G^{\alpha\mathbf{k}+\mathbf{q}}_{\mathbf{r};\lambda}(\omega)
\triangle\Gamma^{\alpha\mathbf{k}+\mathbf{q}}_{\lambda;\mathbf{r}'}(s\mathbf{q};\omega;\nu)\Big\}\widetilde{W}^{\mathbf{q}}_{s;\mathbf{r}}(\nu)
\nonumber\\&=\sum_{\mathbf{q}}w_{\mathbf{q}} \sum_{s}\Big\{A1^{\alpha\mathbf{k}}_{\mathbf{r};\mathbf{r}'}(s\mathbf{q};\tau;\nu)
+e^{-i\nu\tau}A2^{\alpha\mathbf{k}}_{\mathbf{r};\mathbf{r}'}(s\mathbf{q};\tau;\nu)\Big\}
\widetilde{W}^{\alpha\mathbf{q}}_{s;\mathbf{r}}(\nu)\nonumber\\&=\sum_{\mathbf{q}}w_{\mathbf{q}}
\sum_{s}\Big\{B1^{\alpha\mathbf{k}}_{\mathbf{r};\mathbf{r}'}(s\mathbf{q};\tau;\nu)
+e^{-i\nu\tau}B2^{\alpha\mathbf{k}}_{\mathbf{r};\mathbf{r}'}(s\mathbf{q};\tau;\nu)\Big\}\nonumber\\&=
C1^{\alpha\mathbf{k}}_{\mathbf{r};\mathbf{r}'}(\tau;\nu) +e^{-i\nu\tau}C2^{\alpha\mathbf{k}}_{\mathbf{r};\mathbf{r}'}(\tau;\nu).
\end{align}

After that one has generally

\begin{align}\label{sig_9}
\triangle\widetilde{\Sigma}^{dyn,\alpha\mathbf{k}}_{12}(\tau)&=-\frac{1}{\beta}\sum_{\nu}\Big\{e^{i\nu\tau}C1^{\alpha\mathbf{k}}_{12}(\tau;\nu)
+C2^{\alpha\mathbf{k}}_{12}(\tau;\nu)\Big\}\nonumber\\&=-\frac{1}{\beta}\sum_{\nu\geqslant0}\Big\{\cos(\nu\tau)\big\{C1^{\alpha\mathbf{k}}_{12}(\tau;\nu)
+C1^{^{*}\alpha,-\mathbf{k}}_{12}(\tau;\nu)\big\}\nonumber\\&+i\sin(\nu\tau)\big\{C1^{\alpha\mathbf{k}}_{12}(\tau;\nu)
-C1^{^{*}\alpha,-\mathbf{k}}_{12}(\tau;\nu)\big\}\nonumber\\&+C2^{\alpha\mathbf{k}}_{12}(\tau;\nu)
+C2^{^{*}\alpha,-\mathbf{k}}_{12}(\tau;\nu)\Big\}.
\end{align}

\subsection{Correction to the Semi-Dynamic Self Energy} \label{Sig_corr_semi}

Semi-dynamic part of the self energy is divided as the following

\begin{align}\label{sig_10}
&\triangle\widetilde{\Sigma}^{semi}=G\Big\{\triangle\Gamma^{dyn}V+\triangle\Gamma^{stat}\widetilde{W}\Big\}\nonumber\\&=G\Big\{
\triangle\Gamma^{dyn}_{1}V+\triangle\Gamma^{stat}_{1}\widetilde{W}+\triangle\Gamma^{dyn}_{\geqslant2}V+\triangle\Gamma^{stat}_{\geqslant2}\widetilde{W}\Big\},
\end{align}
where the vertex function was divided into the first order and the higher orders.

In the above expression the term $G\triangle\Gamma^{dyn}_{1}V$ is just the transposed of the term $G\triangle\Gamma^{stat}_{1}\widetilde{W}$, so one needs to
calculate only the term $\triangle\widehat{\Sigma}=G\triangle\Gamma^{stat}_{1}\widetilde{W}$:

\begin{align}\label{sig_11}
\triangle\widehat{\Sigma}^{\alpha\mathbf{k}}_{12}(\tau)=&-\sum_{\mathbf{q}}w_{\mathbf{q}}
\sum_{\lambda}G^{\alpha\mathbf{k}+\mathbf{q}}_{1;\lambda}(\tau)\nonumber\\&\times\frac{1}{\beta}\sum_{\nu}e^{i\nu\tau}\sum_{s}
\triangle\Gamma^{\alpha\mathbf{k}+\mathbf{q}}_{\lambda;2}(s\mathbf{q};\nu) \widetilde{W}^{\mathbf{q}}_{s;1}(\nu).
\end{align}

For the different locations of arguments $1$ and $2$ one gets:

Mt-Mt
\begin{align}\label{sig_12}
&\triangle\widehat{\Sigma}^{\alpha\mathbf{k}}_{\mathbf{t}L;\mathbf{t}'L'}(\tau)=-\sum_{\mathbf{q}}w_{\mathbf{q}} \sum_{L''}\sum_{\lambda}
G^{\alpha\mathbf{k}+\mathbf{q}}_{\mathbf{t}L'';\lambda}(\tau)\frac{1}{\beta}\sum_{\nu}e^{i\nu\tau}
\sum_{s}\nonumber\\&\times\triangle\Gamma^{^{stat}\alpha\mathbf{k}+\mathbf{q}}_{\lambda;\mathbf{t}'L'}(s\mathbf{q};\nu) \sum_{s'}\widetilde{W}^{\mathbf{q}}_{s;\mathbf{t}s'}(\nu)
\langle\phi^{\alpha\mathbf{t}}_{L''}|\phi^{\alpha\mathbf{t}}_{L}\Pi^{\mathbf{t}}_{s'}\rangle^{*}\nonumber\\&=-\sum_{\mathbf{q}}w_{\mathbf{q}}
\sum_{L''}\sum_{\lambda} G^{\alpha\mathbf{k}+\mathbf{q}}_{\mathbf{t}L'';\lambda}(\tau)\nonumber\\&\times\frac{1}{\beta}\sum_{\nu}e^{i\nu\tau}
\sum_{s}\triangle\Gamma^{^{stat}\alpha\mathbf{k}+\mathbf{q}}_{\lambda;\mathbf{t}'L'}(s\mathbf{q};\nu) \widetilde{W}^{\alpha\mathbf{q}}_{s;\mathbf{t}L''L}(\nu),
\end{align}

Int-Mt
\begin{align}\label{sig_13}
&\triangle\widehat{\Sigma}^{\alpha\mathbf{k}}_{\mathbf{r};\mathbf{t}'L'}(\tau)=-\sum_{\mathbf{q}}w_{\mathbf{q}} \sum_{\lambda}
G^{\alpha\mathbf{k}+\mathbf{q}}_{\mathbf{r};\lambda}(\tau)\nonumber\\&\times\frac{1}{\beta}\sum_{\nu}e^{i\nu\tau}
\sum_{s}\triangle\Gamma^{^{stat}\alpha\mathbf{k}+\mathbf{q}}_{\lambda;\mathbf{t}'L'}(s\mathbf{q};\nu) \widetilde{W}^{\mathbf{q}}_{s;\mathbf{r}}(\nu),
\end{align}

Mt-Int
\begin{align}\label{sig_14}
&\triangle\widehat{\Sigma}^{\alpha\mathbf{k}}_{\mathbf{t}L;\mathbf{r}'}(\tau)=-\sum_{\mathbf{q}}w_{\mathbf{q}} \sum_{L''}\sum_{\lambda}
G^{\alpha\mathbf{k}+\mathbf{q}}_{\mathbf{t}L'';\lambda}(\tau)\nonumber\\&\times\frac{1}{\beta}\sum_{\nu}e^{i\nu\tau}
\sum_{s}\triangle\Gamma^{^{stat}\alpha\mathbf{k}+\mathbf{q}}_{\lambda;\mathbf{r}'}(s\mathbf{q};\nu) \widetilde{W}^{\alpha\mathbf{q}}_{s;\mathbf{t}L''L}(\nu),
\end{align}

Int-Int
\begin{align}\label{sig_15}
&\triangle\widehat{\Sigma}^{\alpha\mathbf{k}}_{\mathbf{r};\mathbf{r}'}(\tau)=-\sum_{\mathbf{q}}w_{\mathbf{q}} \sum_{\lambda}
G^{\alpha\mathbf{k}+\mathbf{q}}_{\mathbf{r};\lambda}(\tau)\nonumber\\&\times\frac{1}{\beta}\sum_{\nu}e^{i\nu\tau}
\sum_{s}\triangle\Gamma^{^{stat}\alpha\mathbf{k}+\mathbf{q}}_{\lambda;\mathbf{r}'}(s\mathbf{q};\nu) \widetilde{W}^{\mathbf{q}}_{s;\mathbf{r}}(\nu).
\end{align}

The term $G\triangle\Gamma^{dyn}_{\geqslant2}V$ in (\ref{sig_10}) is calculated using (\ref{sig_5}-\ref{sig_8}) similarly to the totally dynamical part but with
$V$ instead of $\widetilde{W}$.

The term $G\triangle\Gamma^{stat}_{\geqslant2}\widetilde{W}$ in (\ref{sig_10}) is calculated using (\ref{sig_12}-\ref{sig_15}) similarly to the semi-dynamical
part of the first order.

\subsection{Correction to the Static Self Energy} \label{Sig_corr_stat}

Totally static part of the self energy is evaluated as the following

\begin{align}\label{sig_16}
\triangle\widetilde{\Sigma}^{stat,\alpha\mathbf{k}}_{12}(\tau)=&-\sum_{\mathbf{q}}w_{\mathbf{q}}
\sum_{\lambda}G^{\alpha\mathbf{k}+\mathbf{q}}_{1;\lambda}(\tau)\nonumber\\&\times\sum_{s} \triangle\Gamma^{^{stat}\alpha\mathbf{k}+\mathbf{q}}_{\lambda;2}(s\mathbf{q};-\tau)
V^{\mathbf{q}}_{s;1}.
\end{align}

Again, for the different locations of arguments $1$ and $2$ one gets:

Mt-Mt
\begin{align}\label{sig_17}
&\triangle\widetilde{\Sigma}^{stat,\alpha\mathbf{k}}_{\mathbf{t}L;\mathbf{t}'L'}(\tau)=-\sum_{\mathbf{q}}w_{\mathbf{q}} \sum_{L''}\sum_{\lambda}
G^{\alpha\mathbf{k}+\mathbf{q}}_{\mathbf{t}L'';\lambda}(\tau)\nonumber\\&\times
\sum_{s}\triangle\Gamma^{^{stat}\alpha\mathbf{k}+\mathbf{q}}_{\lambda;\mathbf{t}'L'}(s\mathbf{q};-\tau) \sum_{s'}V^{\mathbf{q}}_{s;\mathbf{t}s'}
\langle\phi^{\alpha\mathbf{t}}_{L''}|\phi^{\alpha\mathbf{t}}_{L}\Pi^{\mathbf{t}}_{s'}\rangle^{*}\nonumber\\&=-\sum_{\mathbf{q}}w_{\mathbf{q}}
\sum_{L''}\sum_{\lambda} G^{\alpha\mathbf{k}+\mathbf{q}}_{\mathbf{t}L'';\lambda}(\tau)\nonumber\\&\times
\sum_{s}\triangle\Gamma^{^{stat}\alpha\mathbf{k}+\mathbf{q}}_{\lambda;\mathbf{t}'L'}(s\mathbf{q};-\tau) V^{\alpha\mathbf{q}}_{s;\mathbf{t}L''L},
\end{align}

Int-Mt
\begin{align}\label{sig_18}
\triangle\widetilde{\Sigma}^{stat,\alpha\mathbf{k}}_{\mathbf{r};\mathbf{t}'L'}(\tau)=&-\sum_{\mathbf{q}}w_{\mathbf{q}} \sum_{\lambda}
G^{\alpha\mathbf{k}+\mathbf{q}}_{\mathbf{r};\lambda}(\tau) \nonumber\\&\times\sum_{s}\triangle\Gamma^{^{stat}\alpha\mathbf{k}+\mathbf{q}}_{\lambda;\mathbf{t}'L'}(s\mathbf{q};-\tau)
V^{\mathbf{q}}_{s;\mathbf{r}},
\end{align}

Mt-Int
\begin{align}\label{sig_19}
\triangle\widetilde{\Sigma}^{stat,\alpha\mathbf{k}}_{\mathbf{t}L;\mathbf{r}'}(\tau)&=-\sum_{\mathbf{q}}w_{\mathbf{q}} \sum_{L''}\sum_{\lambda}
G^{\alpha\mathbf{k}+\mathbf{q}}_{\mathbf{t}L'';\lambda}(\tau)\nonumber\\&\times
\sum_{s}\triangle\Gamma^{^{stat}\alpha\mathbf{k}+\mathbf{q}}_{\lambda;\mathbf{r}'}(s\mathbf{q};-\tau) V^{\alpha\mathbf{q}}_{s;\mathbf{t}L''L},
\end{align}

Int-Int
\begin{align}\label{sig_20}
\triangle\widetilde{\Sigma}^{stat,\alpha\mathbf{k}}_{\mathbf{r};\mathbf{r}'}(\tau)=&-\sum_{\mathbf{q}}w_{\mathbf{q}} \sum_{\lambda}
G^{\alpha\mathbf{k}+\mathbf{q}}_{\mathbf{r};\lambda}(\tau) \nonumber\\&\times\sum_{s}\triangle\Gamma^{^{stat}\alpha\mathbf{k}+\mathbf{q}}_{\lambda;\mathbf{r}'}(s\mathbf{q};-\tau)
V^{\mathbf{q}}_{s;\mathbf{r}}.
\end{align}

\subsection{Static vertex of the first order} \label{Vrt_1_stat}

Vertex of the first order as a function of $\tau$ ($\triangle\Gamma^{stat}_{1}(\tau)$) should be calculated independently, because, as a function of $\nu$, it is a slow decreasing function, and direct transform $\triangle\Gamma^{stat}_{1}(\tau)=\frac{1}{\beta}\sum_{\nu}e^{-i\nu\tau}\triangle\Gamma^{stat}_{1}(\nu)$ is not easy. Corresponding formulae are obtained straightforwardly:

Mt-Mt
\begin{align}\label{vrt_1_1}
&\triangle\Gamma^{^{stat}\alpha\mathbf{k}}_{1,\mathbf{t}L;\mathbf{t}'L'}(s\mathbf{q};\tau)=\sum_{\mathbf{R}}e^{-i\mathbf{k}\mathbf{R}}
\sum_{L''L'''}V^{\alpha\alpha\mathbf{R}}_{\mathbf{t}LL'';\mathbf{t}'L'L'''}\nonumber\\&\times\frac{1}{N_{\mathbf{k}}}\sum_{\mathbf{k}'}e^{i\mathbf{k}'\mathbf{R}}
K^{0\alpha\mathbf{k}'}_{\mathbf{t}L'';\mathbf{t}'L'''}(s\mathbf{q};\tau),
\end{align}

Int-Mt
\begin{align}\label{vrt_1_2}
\triangle\Gamma^{^{stat}\alpha\mathbf{k}}_{1,\mathbf{r};\mathbf{t}'L'}(s\mathbf{q};\tau)&=\sum_{\mathbf{R}}e^{-i\mathbf{k}\mathbf{R}}
\sum_{L'''}V^{\alpha\mathbf{R}}_{\mathbf{r};\mathbf{t}'L'L'''}\nonumber\\&\times\frac{1}{N_{\mathbf{k}}}\sum_{\mathbf{k}'}e^{i\mathbf{k}'\mathbf{R}}
K^{0\alpha\mathbf{k}'}_{\mathbf{r};\mathbf{t}'L'''}(s\mathbf{q};\tau),
\end{align}

Mt-Int
\begin{align}\label{vrt_1_3}
\triangle\Gamma^{^{stat}\alpha\mathbf{k}}_{1,\mathbf{t}L;\mathbf{r}'}(s\mathbf{q};\tau)&=\sum_{\mathbf{R}}e^{-i\mathbf{k}\mathbf{R}}
\sum_{L''}V^{\alpha\mathbf{R}}_{\mathbf{t}LL'';\mathbf{r}'}\nonumber\\&\times\frac{1}{N_{\mathbf{k}}}\sum_{\mathbf{k}'}e^{i\mathbf{k}'\mathbf{R}}
K^{0\alpha\mathbf{k}'}_{\mathbf{t}L'';\mathbf{r}'}(s\mathbf{q};\tau),
\end{align}

Int-Int
\begin{align}\label{vrt_1_4}
\triangle\Gamma^{^{stat}\alpha\mathbf{k}}_{1,\mathbf{r};\mathbf{r}'}(s\mathbf{q};\tau)&=\sum_{\mathbf{R}}e^{-i\mathbf{k}\mathbf{R}}
V^{\mathbf{R}}_{\mathbf{r};\mathbf{r}'}\nonumber\\&\times\frac{1}{N_{\mathbf{k}}}\sum_{\mathbf{k}'}e^{i\mathbf{k}'\mathbf{R}}
K^{0\alpha\mathbf{k}'}_{\mathbf{r};\mathbf{r}'}(s\mathbf{q};\tau),
\end{align}
with

\begin{align}\label{vrt_1_5}
&K^{0\alpha\mathbf{k}}_{\lambda\lambda'}(s\mathbf{q};\tau)= \nonumber\\&\sum_{\lambda''\lambda'''}G^{\alpha\mathbf{k}}_{\lambda\lambda''}(\tau)
\langle\Psi^{\alpha\mathbf{k}}_{\lambda''}|\Psi^{\alpha\mathbf{k}-\mathbf{q}}_{\lambda'''}\Pi^{\mathbf{q}}_{s}\rangle
G^{\alpha,\mathbf{k}-\mathbf{q}}_{\lambda'''\lambda'}(\beta-\tau).
\end{align}

\section{Definitions of the Matsubara time-frequency transforms} \label{fr_t_2}

For convenience, the definitions of the time-frequency transforms for the functions of two imaginary time arguments, accepted in this work, are collected below.

One starts with general transformations

\begin{align}\label{ww_1}
K(\omega;\omega')&=\int\int d\tau d\tau'e^{i\omega\tau}e^{-i\omega'\tau'} K(\tau;\tau')\nonumber\\&=\int\int d\tau
d\tau'e^{i\omega\tau}e^{i(\omega-\omega')\tau'} K(\tau+\tau';\tau').
\end{align}

Introducing $\nu=\omega-\omega'$ one has

\begin{align}\label{ww_2}
K(\omega;\omega-\nu)&=K(\omega;\nu)\nonumber\\&=\int\int d\tau d\tau'e^{i\omega\tau}e^{i\nu\tau'} K(\tau+\tau';\tau').
\end{align}

From (\ref{ww_2}) other relations follow:

\begin{align}\label{ww_3}
K(\tau;\nu)=\int d\tau'e^{i\nu\tau'} K(\tau+\tau';\tau'),
\end{align}

\begin{align}\label{ww_4}
K(\omega;\nu)=\int d\tau e^{i\omega\tau'} K(\tau;\nu),
\end{align}

\begin{align}\label{tt_1}
K(\tau;\nu)=\frac{1}{\beta}\sum_{\omega}e^{-i\omega\tau}K(\omega;\nu).
\end{align}


\begin{thebibliography}{111}
\bibitem[{k_v()}]{k_vrt_msh}
\bibinfo{note}{This opportunity is under development now. Preliminary tests
  show that k-meshes $2\times2\times2$ or $3\times3\times3$ are already good
  enough for the vertex part in many cases}.
\end{thebibliography}

\end{document}